\documentclass[aps,preprint,superscriptaddress,groupedaddress,nofootinbib]{revtex4}        

\usepackage{verbatim}
\usepackage{multirow}
\usepackage{dcolumn}   
\usepackage{bm}        
\usepackage{amssymb}   
\usepackage{amsmath}
\usepackage{amsfonts}
\usepackage{graphicx}
\usepackage[ngerman, english]{babel}
\usepackage{float}
\usepackage{subfigure}
\usepackage[utf8]{inputenc}
\usepackage{mathrsfs}
\usepackage{hhline}
\usepackage{braket}
\usepackage{bbold}
\usepackage{nicefrac}
\usepackage{xcolor}
\usepackage{hyperref}
\usepackage{natbib}
\usepackage{comment}
\usepackage{pifont}
\usepackage{footmisc}
\usepackage[section]{placeins}

\def\block(#1,#2)#3{\multicolumn{#2}{c}{\multirow{#1}{*}{$ #3 $}}}

\renewcommand {\t}{\tilde{\theta}}

\newcommand{\cmark}{\textcolor{green}{\ding{51}}}
\newcommand{\xmark}{\textcolor{red}{\ding{55}}}

\begin{document}

\title{Sterile Neutrinos with Altered Dispersion Relations as an Explanation for Neutrino Anomalies}

 \preprint{DO-TH 18/17}
 \author{Dominik D\"oring}
 \email[]{dominik.doering@tu-dortmund.de}

 \author{Heinrich P\"as}
 \email[]{heinrich.paes@tu-dortmund.de}
 
 \author{Philipp Sicking}
 \email[]{philipp.sicking@tu-dortmund.de}

 \affiliation{Fakult\"at f\"ur Physik,
	 Technische Universit\"at Dortmund, 44221 Dortmund,
 Germany}

\author{Thomas J. Weiler}
\email[]{tom.weiler@vanderbilt.edu}
\affiliation{Department of Physics \& Astronomy, Vanderbilt University, Nashville TN 37235, USA}

\begin{abstract}
    Recently the MiniBooNE Collaboration has confirmed its anomalous excess in
    $\overset{\scriptscriptstyle(-)}{\nu}_\mu \rightarrow 
    \overset{\scriptscriptstyle(-)}{\nu}_e$ neutrino oscillation data. Combined with long-standing results from the
    LSND experiment this amounts to a $6.1\,\sigma$ evidence for new physics beyond the Standard Model. In this paper
    we develop a framework with 3 active and 3 sterile neutrinos with altered dispersion relations that provides a mechanism to explain these
    anomalies without being in conflict with the absence of anomalous neutrino disappearance in other neutrino oscillation experiments.
\end{abstract}

\maketitle

 \section{Introduction}
 The recently reported results of the MiniBooNE collaboration \cite{Aguilar-Arevalo:2018gpe, Aguilar-Arevalo:2020nvw} exhibit a $4.8\,\sigma$ excess for $
\overset{\scriptscriptstyle(-)}{\nu}_\mu \rightarrow 
\overset{\scriptscriptstyle(-)}{\nu}_e$ transitions
 which cannot be explained by the standard three neutrino picture. Combined with the earlier LSND results \cite{Aguilar:2001ty}, 
 these findings even provide a significance of $6.1\, \sigma$ for new physics beyond the Standard Model. 
The high significance demonstrates that this excess is not simply due
 to statistics, but has to be systematic.
 Albeit it is possible that the excess is due to an underestimation of the background, it is worthwhile to explore the possibility of 
 this excess being a sign for new physics in the neutrino oscillation regime. Moreover, several other anomalies, like the
 the Reactor-~\cite{Mention:2011rk} and Gallium anomalies~\cite{Giunti:2010zu} also provide hints for
 new physics or additional sterile neutrinos. At the same time atmospheric neutrino experiments~\cite{TheIceCube:2016oqi, Abe:2014gda, Aartsen:2020iky}
 or accelerator experiments~\cite{Abe:2014nuo, Adamson:2017uda, Adamson:2017zcg}
 searching for $\nu_\mu$ disappearance do not
 show any deviation from the standard three neutrino picture and yield stringent constraints on additional sterile neutrinos that rule out the most simple models. 
 
More concretely, neutrino data exhibit:

\begin{itemize}

\item An excess of data, interpretable as $\overset{\scriptscriptstyle(-)}{\nu}_\mu \rightarrow \overset{\scriptscriptstyle(-)}{\nu}_e$ 
oscillations in both the neutrino and the anti-neutrino channels in the LSND and MiniBooNE data.
\item A resonance-like excess in MiniBooNE's low energy data. 
\item Evidence for electron neutrino disappearance in experiments utilizing a Gallium source or reactor neutrinos.
\item No anomalous neutrino or anti-neutrino disappearance at higher energy accelerator experiments or atmospheric neutrino experiments.
\end{itemize}

In this paper we discuss the
problems associated with the conflicting results from appearance and disappearance experiments 
(for a recent review see \cite{Dentler:2018sju})
and develop a simple model adopting altered dispersion relations (ADRs) for sterile neutrinos which 
may explain both results consistently and describe the world's neutrino data successfully. 
Other model set-ups, which can successfully reconcile LSND and MiniBooNE with disappearance results are e.g. decaying neutrino models \cite{deGouvea:2019qre, Dentler:2019dhz}, while \cite{deGouvea:2019qre} does not aim to explain the reactor anomaly at the same time. There are also models that seek to resolve these anomalies with the help of sterile neutrinos in Large Extra Dimensions (LED), e.g. \cite{Carena:2017qhd}.

The paper is organized as follows: In section 2 we review the general framework of neutrino oscillations in the presence of sterile neutrinos. 
Section 3 discusses the effect of altered dispersion relations and points out the phenomenological difficulties that arise when only a single sterile 
neutrino is assumed. 
Section 4 discusses the complete $3+3\nu$ neutrino model and develops a phenomenological framework that could be able to successfully describe the present neutrino data. Phenomenological implications of such models and several benchmark points in the parameter space are discussed in section 5. Finally, sections 6 and 7 briefly address open questions and cosmological constraints.

\section{Neutrino Oscillations in the Presence of Sterile Neutrinos}

 The transition probability of neutrinos from one active flavor $\alpha$ to another flavor $\beta$ can be written as
 \begin{align}
	 P_{\nu_\alpha\rightarrow\nu_\beta} 
	 =& \delta_{\alpha\beta} 
	 -4\sum_{k>j}^{N} \operatorname{Re}( U_{\alpha k}^*U_{\beta k}^{\ } U_{\alpha j}^{\ } U_{\beta j}^*)
	\sin^2{\left(\frac{\Delta m_{kj}^2L}{4E}\right)}\nonumber \\
	&+{2\sum_{k>j}^{N}\ \operatorname{Im}(U_{\alpha k}^*U_{\beta k}^{\ } U_{\alpha j}^{\ } U_{\beta j}^*)
	\sin{\left(\frac{\Delta m_{kj}^2L}{2E}\right)}}\text{,}
	\label{eq:basicOscillation}
 \end{align}
 with $N$ being the total number of active and sterile light neutrinos, $U$ the $N\times N$ unitary mixing-matrix, 
 $\Delta m_{kj}^2= m_k^2-m_j^2$ the mass-squared differences of the mass eigenvalues, $L$ the baseline and $E$ the energy of the 
 neutrino. From here on we neglect possible CP violation for all practical purposes and consider only real elements of the 
 mixing matrix $U$, so the last term in Eq. \eqref{eq:basicOscillation} vanishes.  Also, the CPT theorem and CP invariance imply that 
 T is a good symmetry, so $\sin^2{2\theta_{\alpha\beta}}=\sin^2{2\theta_{\beta\alpha}}$ pertains.
Equation~\eqref{eq:basicOscillation} holds for mixing with any number of additional sterile neutrinos.

 As the proposed additional mass-squared difference $\Delta m^2_\text{LSND}$ is in the $\sim 1\, \text{eV}^2$-region and the 
 mass-squared differences $\Delta m_{21}^2$, $\Delta m_{31}^2$ and $\Delta m_{32}^2$ are experimentally tested to be orders of
 magnitude smaller, it is possible to neglect 
 $\Delta m^2_{21}$, $\Delta m_{31}^2$ and $\Delta m_{32}^2$,
 Therefore the transition probability for sensible values of $\nicefrac{L}{E}$ and a 3+1$\nu$ model reduces to
 \begin{align}
	 P_{\nu_\alpha\rightarrow\nu_\beta} 
	 =& \delta_{\alpha\beta} 
	-4\sum_{j}^{3} \operatorname{Re}( U_{\alpha 4}^*U_{\beta 4}^{\ } U_{\alpha j}^{\ } U_{\beta j}^*)
	\sin^2{\left(\frac{\Delta m_{4j}^2L}{4E}\right)}\nonumber\\
	=& \delta_{\alpha\beta} - 4 \sin^2{\left(\frac{\Delta m_{\mathrm{LSND}}^2L}{4E}\right)}
	\underbrace{\sum_{j}^{3} U_{\alpha 4}U_{\beta 4} U_{\alpha j} U_{\beta j}}_{-\frac{1}{4}\sin^2{2\theta_{\alpha\beta}}\,,
	{\rm \ for\ }\alpha \neq \beta}
	\, \text{.} 
	 \label{eq:4nuOscillation}
 \end{align}
Due to unitarity of $U$, one has 
$\sum_j^3 U_{\alpha j} U_{\beta j}= - U_{\alpha 4} U_{\beta 4}+\delta_{\alpha\beta}$,
and the resulting effective amplitudes can be reduced to the appearance value 
$\sin^2{2\theta_{\alpha\beta}}= 4 |U_{\alpha 4}|^2 |U_{\beta 4}|^2$ 
and the disappearance value $\sin^2{2\theta_{\alpha\alpha}}= 4 |U_{\alpha 4}|^2 (1-|U_{\alpha 4}|^2)$.
It is obvious that at the appropriate mass, energy and baselength scales a $3+1\nu$ model with a $\sim 1 \mathrm{eV}^2$ mass
 squared difference looks similar to, and can be analyzed like, a two neutrino scenario. 

 Appearance experiments like LSND or MiniBooNE measure the transition $\overset{\scriptscriptstyle(-)}{\nu}_\mu \to \overset{\scriptscriptstyle(-)}{\nu}_e$ 
 and therefore require a sizable value for $\sin^2 2\theta_{\mu e}$, in order to explain the observed excess.
 On the other hand, disappearance experiments do not observe a significant deficit in 
 $\overset{\scriptscriptstyle(-)}{\nu}_\mu \to \overset{\scriptscriptstyle(-)}{\nu}_\mu$ 
or $\overset{\scriptscriptstyle(-)}{\nu}_e \to \overset{\scriptscriptstyle(-)}{\nu}_e$ 
 oscillations, and therefore constrain the corresponding value for 
 $\sin^2{2\theta_{\mu\mu}}=4\,|U_{\mu4}|^2(1-|U_{\mu4}|^2)\simeq 4\,|U_{\mu4}|^2$ for $|U_{\mu4}|^2$~small, and 
 $\sin^2{2\theta_{ee}}=4\,|U_{e4}|^2(1-|U_{e4}|^2)\simeq 4\,|U_{e4}|^2$ for $|U_{e4}|^2$~small \footnote{The survival amplitude can also be written as 
 $\sin^2 2\theta_{\alpha\alpha} = 1- 4\,(|U_{\alpha\alpha}|^2-\frac{1}{2} )^2$,
  which shows the symmetry about the midpoint $|U_{\alpha\alpha}|^2 = \frac{1}{2}$.
  Note that for a given $\sin^2 2\theta_{\alpha\alpha} < 1$, there are two solutions of $|U_{\alpha\alpha}|^2$,
  $|U_{\alpha\alpha}|^2$ and $1 - |U_{\alpha\alpha}|^2$.
}. 
 By comparing the different amplitudes, the following relation is derived (as in e.g. \cite{Peres:2000ic}):
 
 \begin{align}
	 \sin^2 2\theta_{\mu e}&= 4\, |U_{\mu 4}|^2 |U_{e 4}|^2 = \frac{1}{4} (\sin^2 2\theta_{\mu\mu} +4\,|U_{\mu\mu}|^4) 
	 	(\sin^2 2\theta_{ee} +4\,|U_{ee}|^4)\nonumber\\
	 & \simeq \frac14 \sin^2 2\theta_{\mu\mu}\sin^2 2\theta_{ee}\text{,}
\end{align}
where the exact third expression becomes the approximate result since it is known from data that 
$\sin^2{2\theta_{\mu s}}= 4 |U_{\mu 4}|^2 |U_{s 4}|^2$ and $\sin^2{2\theta_{e s}}= 4 |U_{e 4}|^2 |U_{s 4}|^2$ 
are small.

Both disappearance probabilities ($\overset{\scriptscriptstyle(-)}{\nu}_e$ and $\overset{\scriptscriptstyle(-)}{\nu}_\mu$) have to be relatively large 
 to generate a sufficient $\sin^2 (2\theta_{e\mu})$.  But large values   
 contradict the data from reactor and disappearance accelerator experiments. 
 This relation is well known and exhibits the basic problem of current short baseline results.
 The problem persists in models adopting more than one additional sterile state~\cite{Giunti:2015mwa}.

 The above relationship holds as long as one considers the elements of the mixing matrix to be constant. If this condition is given up
 and the elements are allowed to become energy or baseline dependent (as is the case for CP-violating matter vs.~anti-matter effects), 
 this tension can in principle be avoided.
 The relevant experiments such as MINOS ($E\sim 7$ GeV, $L\sim 735$ km), typical atmospheric neutrino experiments 
 ($E\sim 0.6 - 100$ GeV, $L\sim 15 - 13,000$ km), LSND ($E\sim 10-100$ MeV, $L\sim 30$ m) and 
 MiniBooNE ($E\sim 0.2 - 3$ GeV, $L\sim 541$ m) are indeed all sensitive to a mass-squared difference of $\sim 1\, \mathrm{eV}^2$ due
 to similar values of $\nicefrac{L}{E}$, but they operate on different energy and base-length scales and so an altered dispersion relation 
 could be useful to resolve the above mentioned problem. 
 In this case the energy dependence has to be fairly strong, since the energy regime for MiniBooNE almost
 overlaps with the low-energy range of atmospheric neutrino experiments.

 So to relieve the tension between appearance and disappearance experiments, a model is required 
 which allows for a small $|U_{\mu4}|^2$  at high energies (GeV) and 
 a sufficiently large $|U_{e4}|^2|U_{\mu4}|^2$ at low energies (MeV).

 \section{Altered Dispersion Relations for a Single Sterile Neutrino}
 Scenarios with altered dispersion relations (ADRs) adopt additional terms in the
 usual relation between energy $E$ and momentum $\vec{p}$, so
 $E^2 \neq \left|\vec{p}\right|^2 + m^2$.
 As we demonstrate below, 
 energy dependent elements of the mixing matrix and mass-squared
 differences can be generated by an additional effective potential in the
 Hamiltonian in flavor space.

A typical example for such effects are neutrino matter effects, as they
 arise e.g.~in the MSW description of solar neutrinos
 \cite{Mikheev:1986wj}. Matter effects imply a new, energy-independent term in the
 Hamiltonian and are significant typically either
 for neutrinos or for anti-neutrinos, but not for both. In addition, a solution for the
 MiniBooNE anomaly exploiting matter effects would require unusually
 large couplings. The formalism developed in the following
 can be adapted to include matter effects, but their effects are expected to be small.
 Older data sets were also studied in the light of non-standard matter effects 3+1$\nu$ models indicating positive effects towards a compatibility of LSND and MiniBooNE data \cite{Karagiorgi:2012kw}.

Scenarios with sterile neutrino Lorentz violation allow for different
 energy dependencies which typically apply to neutrinos and
 anti-neutrinos in the same way.
 This is because the application of Lorentz violation affects spacetime directly, 
 and not particles vs.~antiparticles.
 The model proposed in \cite{Pas:2005rb,Marfatia:2011bw} 
 (see also~\cite{Hollenberg:2009bq,Hollenberg:2009ws,Aeikens:2014yga,Aeikens:2016rep})
 adopts one additional sterile neutrino taking a shortcut via an
 asymmetrically warped extra dimension~\cite{Chung:1999zs,Chung:1999xg,Csaki:2000dm}.
In a semi-classical picture, the sterile neutrino oscillates on its geodesic in the warped bulk surrounding the brane, 
and thereby a running time difference is generated between active and sterile neutrinos. 
This running time difference manifests itself
as an additional negative potential in the Hamiltonian proportional to
the relative time difference $\frac{\delta t}{t}=: \varepsilon$, the so-called shortcut parameter 
(always entering the Hamiltonian as multiplied by the energy $E$).
Although the semi-classical picture may not be truly accurate, its predictions regarding
the form of the potential are correct to leading order in the shortcut parameter \cite{Doring:2018ncz}.

This leads us to the assumption that the resulting Hamiltonian \footnote{Standard matter effects can simply be added to this scenario in flavor space in the usual way.} in flavor space can be written as 
\begin{align}
	H_{\mathrm{(F)}}&= \frac{1}{2E} U \begin{pmatrix} m_1^2 & 0 & 0 & 0\\
		0 & m_2^2 & 0 & 0 \\
		0 & 0 &m_3^2 & 0\\
		0 & 0 & 0 & m_4^2 \end{pmatrix}
	U^\dagger - E 
	\begin{pmatrix} 0 & 0 & 0 & 0\\
		0 & 0 & 0 & 0 \\
		0 & 0 &0 & 0\\
		0 & 0 &0 & \varepsilon \end{pmatrix} \nonumber \\
	&\approx\begin{pmatrix} \block(3,3){V}& 0  \\  & & & 0  \\ & & & 0  \\0& 0& 0 & 1\end{pmatrix}
	\left[ \frac{1}{2E} \begin{pmatrix} 1 & 0 & 0 & 0 \\
			0 & 1 & 0 & 0 \\
			0 & 0 & \block(2,2){R_{34}} \\
			0 & 0& & \end{pmatrix}
		\begin{pmatrix} 0 & 0 & 0 & 0\\
			0 & 0 & 0 & 0 \\
			0 & 0 & 0 & 0\\
			0 & 0 & 0 & \Delta m^2_{\mathrm{LSND}} \end{pmatrix}
		\begin{pmatrix} 1 & 0 & 0 & 0 \\
			0 & 1 & 0 & 0 \\
			0 & 0 & \block(2,2){R_{34}^T} \\
			0 & 0& & \end{pmatrix}
		- E\varepsilon
		\begin{pmatrix} 0 & 0 & 0 & 0\\
			0 & 0 & 0 & 0 \\
			0 & 0 & 0 & 0\\
		0 & 0 &0 & 1 \end{pmatrix} \right]
	\begin{pmatrix} \block(3,3){V^\dagger}& 0  \\  & & & 0  \\& & & 0  \\ 0& 0& 0 & 1
	\end{pmatrix} \text{,}
	\label{eq:HamFlavor4nu}
\end{align}
with the energy $E$, shortcut parameter $\varepsilon$, and
\begin{align}
	U = \begin{pmatrix} \block(3,3){V}& 0  \\  & & & 0 \\  & & & 0  \\ 0& 0& 0 & 1\end{pmatrix} \times \begin{pmatrix} 1 & 0 & 0 & 0 \\
		0 & 1 & 0 & 0 \\
		0 & 0 & \block(2,2){R_{34}} \\
		0 & 0& & \end{pmatrix}\, \text{,}
	\label{}
\end{align}
being the full $4\times 4$ unitary mixing matrix. 
Here, $V$ is the unitary $3\times 3$ mixing matrix corresponding to 
the standard $U_{\mathrm{PMNS}}$
and $R_{34}$ is the rotation in the $3-4$ plane generating the sterile admixture of mass eigenstate $\nu_3$ with the mixing angle
$\theta_{34}$:
\begin{align}
	R_{34}=\begin{pmatrix} \cos{\theta_{34}} & \sin{\theta_{34}}\\ -\sin{\theta_{34}} & \cos{\theta_{34}} \end{pmatrix} \text{.} 
	\label{}
\end{align}
In the last line of Eq. \eqref{eq:HamFlavor4nu}, we have used the aforementioned approximation $m_1^2 = m_2^2=m_3^2=0$ and
$m_4^2 = \Delta m^2_{\mathrm{LSND}}$. 

As already calculated in \cite{Marfatia:2011bw}, the eigenvalues of the Hamiltonian become 
\begin{align}
	\lambda_1=\lambda_2=0\mathrm{,} \quad \lambda_{\pm}=\frac{\Delta m^2_{\mathrm{LSND}}}{4E}\left(1-\cos{2\theta_{34}}
	\left(\frac{E}{E_R}\right)^2 \pm \sqrt{\sin^2{2\theta_{34}}+\cos^2{2\theta_{34}}
	\left[1-\left(\frac{E}{E_R}\right)^2\right]^2}\right)\, \text{,}
	\label{}
\end{align}
with the resonance energy 
\begin{align}
\label{E_R}
E_R = \sqrt{\frac{\Delta m^2_{\mathrm{LSND}} \cos{2\theta_{34}}}{2\varepsilon}}\,.
\end{align}
Below we follow the arguments given by~\cite{Marfatia:2011bw} with one 
exception regarding the reasoning why the probability 
$P_{\nu_\alpha \rightarrow\nu_\alpha}$, with $\alpha$ being an active flavor, should vanish.
One has
\begin{align}
	P_{\nu_\alpha \rightarrow\nu_\alpha} = 1-4 U_{\alpha 3}^2 \times 
	\begin{cases} \sin^2{\left(\frac{L(\lambda_+ - \lambda_-)}{2}\right)} 
		\sin^2\t \cos^2\t \, U_{\alpha 3}^2 \\
		\sin^2{\left(\frac{L(\lambda_+)}{2}\right)} \sin^2\t \left(1- U_{\alpha 3}^2\right)\\ 
		\sin^2{\left(\frac{L(\lambda_-)}{2}\right)} \cos^2\t \left(1- U_{\alpha 3}^2\right)\, \text{.}
	\end{cases}
	\label{eq:probab4nu}
\end{align}
Here, $\t$ denotes the effective mixing angle defined via 
\begin{align}
	\sin^2{2\t}=\frac{\sin^2{2\theta_{34}}}{\sin^2{2\theta_{34}}+\left(\cos{2\theta_{34}}-\frac{2E^2\varepsilon}
	{\Delta m^2_{\mathrm{LSND}}}\right)^2} \, \text{.}
\end{align}
According to \cite{Marfatia:2011bw}, while $\sin^2{\t}$ does not vanish, the eigenvalue $\lambda_+$ vanishes and 
therefore $P_{\nu_\alpha \rightarrow\nu_\alpha}$ should vanish as well.
Technically, this is a correct statement which applies as long as one considers only a single experiment with a fixed 
base-length $L$. However, various experiments now observe neutrinos in a wide energy range above the resonance. 
For example, for atmospheric experiments
not only the energy becomes larger than the energies at LSND or MiniBooNE, but also the base-length can be as large as
15,000 km, which results in a value of $\nicefrac{L}{E}$ which is up to 4 magnitudes larger than the one probed 
in MiniBooNE.

Therefore, the relevant quantities to be examined are the mass-squared eigenvalues rather than the Hamiltonian eigenvalues.
The mass-squared eigenvalues are
\begin{align}
	m^2_{\pm}=2E \cdot \lambda_{\pm} \, \text{,}
	\label{}
\end{align}
which give rise to the oscillatory term $\sin^2{\left(m^2_\pm \frac{L}{2E}\right)}$ in the probability in Eq.~\eqref{eq:probab4nu}. 
Adopting this more familiar form we continue to analyze the oscillations of atmospheric neutrinos.
While it is true that $\lambda_+ \propto 1/E$ becomes zero for energies much larger than $E_R$, $m^2_+$ as defined herein does not:
\begin{align}
	\lim_{E\to\infty} m^2_+ = \Delta m^2_{\mathrm{LSND}}\cdot \frac{1-\cos{2\theta_{43}}}{2} \, .
	\label{}
\end{align}
Note that, although the effective mass-squared eigenvalue $m^2_+$ can become small for extremely small mixing angles $\theta_{34}$, the
important value is still the bare (or ``vacuum") mass of the 4th eigenvalue, in this case $\Delta m^2_{\mathrm{LSND}}$. This value is still
large compared to the standard mass-and squared differences.
Therefore, even above the resonance there is still a non-vanishing oscillation mode, which becomes accessible experimentally if 
the oscillation length is large enough.  Such is the case for atmospheric and astrophysical neutrinos.

\begin{figure}[H]
	\centering
	\includegraphics[width=0.79\textwidth]{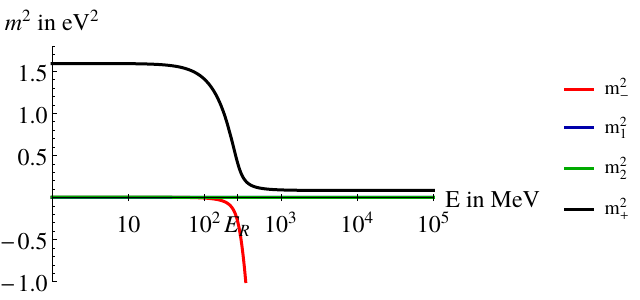}
	\caption{Effective mass-squareds as a function of the energy $E$ in the $3+1\nu$ model including an effective potential due to 
		sterile neutrino ADRs.}
	\label{fig:masses4nu}
\end{figure}

For a better understanding we plot the effective masses for this scenario in Fig.~\ref{fig:masses4nu}.
As can be seen in Fig.~\ref{fig:masses4nu}, at the resonance a level crossing occurs and the Hamiltonian eigenstates swap their
flavor content. While the predominantly sterile state decouples above the resonance, the now heavier predominantly active state
approaches a constant value that due to the level crossing gap is different from its initial value.
This implies a large effective $\Delta m^2_{13}$ that gives rise to large and fast active-to-sterile oscillations, e.g. in atmospheric
neutrinos.

This argument can be generalized to mixing with $\theta_{14}$, $\theta_{24}$, or a combination of the two.
(Also, the case where all standard mass-squared differences do not vanish can be treated accordingly.)

Although the sterile neutrino decouples from the active ones above the resonance, 
the impact on the disappearance experiments is significant. As has been first discovered numerically by Patrick Huber, 
in any possible mixing pattern the atmospheric experiments or MINOS should notice
a deviation from the standard three neutrino case (especially for longer baselines such as the MINOS far detector, or upward going atmospheric 
neutrinos). One could possibly argue that MINOS might miss the deviation from standard 3 neutrino case due to its narrow energy 
spectrum at around $7$ GeV, but atmospheric experiments like IceCube or SuperKamiokaNDE or KM3NeT have not only high statistics but also a 
wide energy spectrum and high resolution for the azimuth angle and the energy. 
Therefore, these experiments should be highly sensitive to this deviation. 
Atmospheric neutrino experiments also tested the $\nicefrac{L}{E}$-dependence 
of the oscillation probability, finding no observable deviations from the $3 \nu$ case.
A simple $3+1\nu$ model even with an altered dispersion relation for the sterile neutrino is therefore ruled out by current data.

\section{A Realistic 3+3 Model}

\subsection{$3+3 \nu$ with a Common Sterile Neutrino Potential}
The emergence of a large mass-squared difference in the energy regime far above the resonance described above can be 
avoided by adding three sterile neutrinos instead of a single sterile neutrino. In the following we assume that 
each sterile state mixes with one of the predominantly active mass eigenstates, respectively. 
Therefore all mass eigenstates become affected by the common sterile potential. If all sterile neutrinos
are affected by the same potentials, the mass differences among the predominantly active states will not be altered even though their 
masses change as the $E$-dependent potential and mixings change.
This mechanism removes the `unwanted' mass difference which spoiled atmospheric neutrino oscillations in the $3+1\nu$ case.

The resulting $6\times 6$ mixing matrix can be parametrized as
\begin{align}
	U^{6\times 6} = U_{23}U_{13}U_{12}U_{14}U_{25}U_{36}\, \text{.}
	\label{}
\end{align}
The bare masses read 
\begin{align}
	\Delta m_{41}^2 &= \Delta m^2_{\mathrm{LSND}}\, \text{,}\\
	\Delta m_{51}^2 &= \Delta m^2_{\mathrm{LSND}}+\Delta m^2_{21} \quad \rightarrow \quad \Delta m^2_{52}=\Delta m^2_{\mathrm{LSND}}\, \text{,}\\
	\Delta m_{61}^2 &= \Delta m^2_{\mathrm{LSND}}+\Delta m^2_{31} \quad \rightarrow \quad \Delta m^2_{63}=\Delta m^2_{\mathrm{LSND}}\, \text{.}
	\label{}
\end{align}
Assuming universal resonance energies ($E_R$) or ADR parameters ($\varepsilon$) for the three sterile neutrinos,
the effective potential that results is
\begin{align}
	V_{\mathrm{eff}}= \begin{pmatrix} 0&0&0&0&0&0\\ 
		0&0&0&0&0&0\\ 
		0&0&0&0&0&0\\ 
		0&0&0&\varepsilon E&0&0\\
		0&0&0&0&\varepsilon E &0\\
		0&0&0&0&0&\varepsilon E\end{pmatrix}\text{.}
	\label{}
\end{align}
Every mass eigenstate $\nu_{1,2,3}$ has its own sterile state admixture.  Still, the results from \cite{Marfatia:2011bw} remain applicable.  
The resulting mass eigenvalues are denoted by $m_{4\pm}^2$, $m_{5\pm}^2$, $m_{6\pm}^2$, which correspond to the $m^2_\pm$ mass
eigenstates in the previous section. The eigenvalues
read 
\begin{align}
	 m^2_{4\pm}\approx  \frac{\Delta m^2_{\mathrm{LSND}}}{2}\left(1 - \cos{2\theta_{14}}
	\left(\frac{E}{E_{R,4}}\right)^2 \mp \sqrt{\sin^2{2\theta_{14}}+\cos^2{2\theta_{14}}
	\left[1-\left(\frac{E}{E_{R,4}}\right)^2\right]^2}\right)\label{eq:masses1}\, \text{,}\\
	 m^2_{5\pm}\approx  \frac{\Delta m^2_{\mathrm{LSND}}}{2}\left(1 - \cos{2\theta_{25}}
	\left(\frac{E}{E_{R,5}}\right)^2 \mp \sqrt{\sin^2{2\theta_{25}}+\cos^2{2\theta_{25}}
	\left[1-\left(\frac{E}{E_{R,5}}\right)^2\right]^2}\right)\label{eq:masses2}\, \text{,}\\
	 m^2_{6\pm}\approx \frac{\Delta m^2_{\mathrm{LSND}}}{2}\left(1 - \cos{2\theta_{36}}
	\left(\frac{E}{E_{R,6}}\right)^2 \mp \sqrt{\sin^2{2\theta_{36}}+\cos^2{2\theta_{36}}
	\left[1-\left(\frac{E}{E_{R,6}}\right)^2\right]^2}\right)\label{eq:masses3}\, \text{,}
\end{align}
with the corresponding resonance energies
\begin{align}
	E_{R,4}=\sqrt{\frac{\Delta m^2_{\mathrm{LSND}} \cos{2\theta_{14}}}{2\varepsilon}}\, \text{,}\quad
	E_{R,5}=\sqrt{\frac{\Delta m^2_{\mathrm{LSND}}\cos{2\theta_{25}}}{2\varepsilon}}\, \text{,}\quad
	E_{R,6}=\sqrt{\frac{\Delta m^2_{\mathrm{LSND}}\cos{2\theta_{36}}}{2\varepsilon}}\, \text{.}
	\label{eq:resonanceEnergies}
\end{align}
All mass-squareds $\Delta m^2_{i-}$ approach minus infinity 
and decouple in the limit $E\gg E_{R,i}$. 
The mass-squared eigenvalues of interest are again $\Delta m^2_{i+}$, whose values in the limit $E\gg E_{R,i}$ are
\begin{align}
	\lim_{E\to\infty}  m^2_{4+}  \approx \frac12 \left(\Delta m^2_{\mathrm{LSND}}\cdot \cos{2\theta_{14}}\right)\, \text{,}\\
	\lim_{E\to\infty} m^2_{5+}   \approx \frac12 \left(\Delta m^2_{\mathrm{LSND}}\cdot \cos{2\theta_{25}}\right)\, \text{,}\\
	\lim_{E\to\infty}  m^2_{6+}  \approx \frac12 \left(\Delta m^2_{\mathrm{LSND}}\cdot \cos{2\theta_{36}}\right)\, \text{,}
	\label{}
\end{align}
while the relevant mass-squared differences far above the resonance become
\begin{align}
	m_{5+}^2-m_{4+}^2  \approx \frac{\Delta m^2_{\mathrm{LSND}}}{2}\left(\cos{2\theta_{14}}-\cos{2\theta_{25}}\right)\, \text{,}
	\label{eq.farAbove1} \\
	m_{6+}^2-m_{4+}^2  \approx \frac{\Delta m^2_{\mathrm{LSND}}}{2}\left(\cos{2\theta_{14}}-\cos{2\theta_{36}}\right)\, \text{,}\\
	m_{6+}^2-m_{5+}^2  \approx \frac{\Delta m^2_{\mathrm{LSND}}}{2}\left(\cos{2\theta_{25}}-\cos{2\theta_{36}}\right)\, \text{'}
	\label{eq:farAbove3}
\end{align}
If these mass-squared differences are all assumed to lie in
the same region as the mass-squared difference in the $3+1\nu$ case, a corresponding oscillation
should be measurable in atmospheric neutrino experiments.
Such an ``extra" oscillation is not seen. 

\begin{figure}[h]
	\centering
	\includegraphics[width=0.99\textwidth]{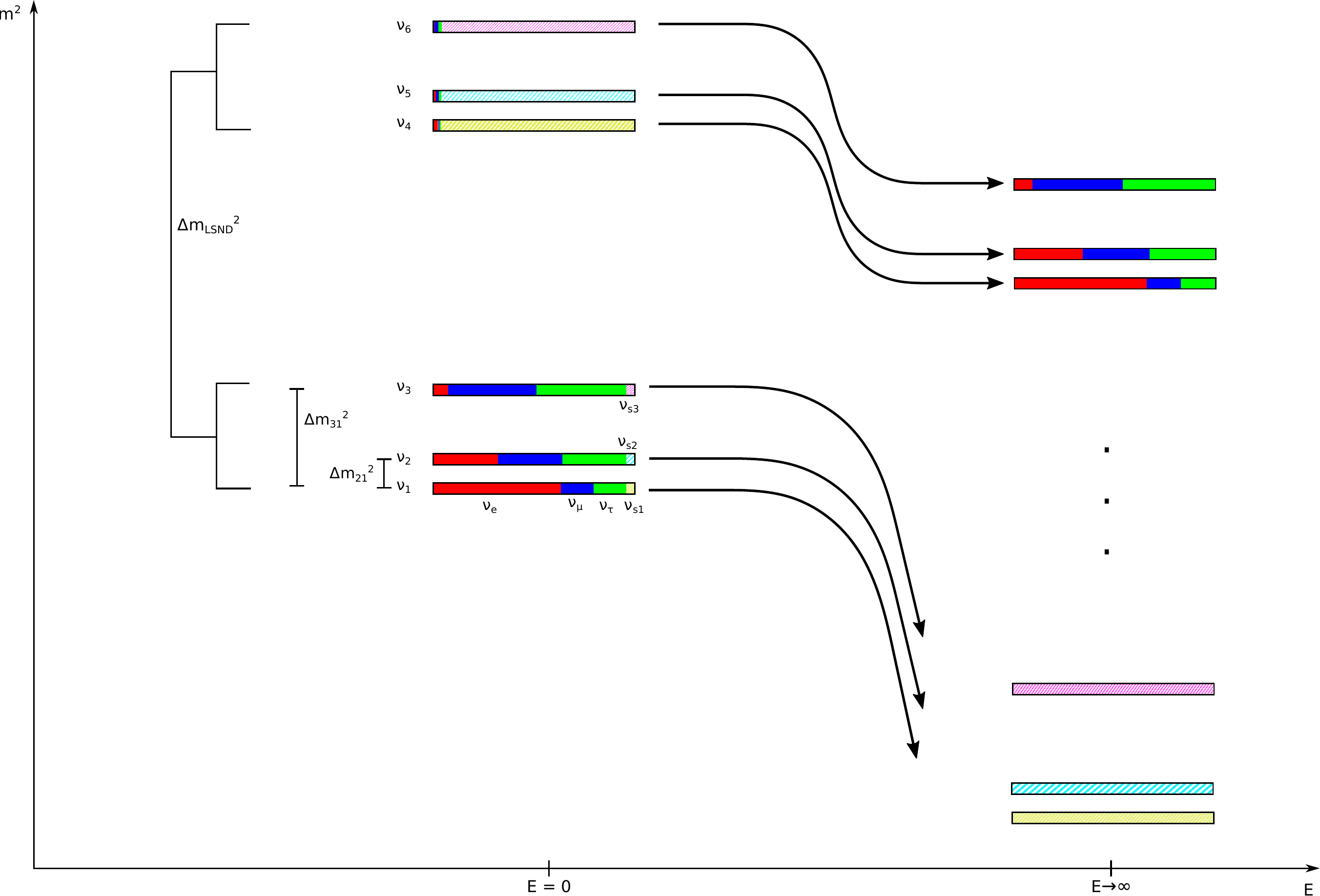}
	\caption{Schematic overview of mass eigenstates and their flavor content depending on the Energy $E$.}
	\label{fig:schematic}
\end{figure}

The only way to avoid the generation of such a mass-squared difference, is by imposing a common mixing in addition to the common potential, 
i.e.\@ by setting all 
new mixing angles to the same value: $\theta_{14}=\theta_{25}=\theta_{36}\equiv \theta$. 
In this case the terms in Eqns.~\eqref{eq.farAbove1}-\eqref{eq:farAbove3}
vanish for all mass-squared differences, and by construction of the model one ends up with the same mass-squared differences as in a standard 
three neutrino scenario (see Fig.~\ref{fig:schematic}).
Consequently, it is possible to avoid the constraints by atmospheric neutrino experiments.
However, as long as the resonance energies for the three sterile neutrinos are assumed to be universal, a
new problem arises in the form of a vanishing amplitude 
for the MiniBooNE experiment in the resonant region.

In general, the oscillation probability including the active mass-squared differences, in the case of CP conservation, is given by
\begin{align}
	P_{\nu_\mu\rightarrow\nu_e} = -4\sum_{k>j}\tilde{U}_{\mu j}\tilde{U}_{\mu k}\tilde{U}_{e j}\tilde{U}_{e k} 
	\sin^2{\left(\Delta\tilde{m}_{kj}^2\frac{L}{2E}\right)}\, \text{.}
	\label{}
\end{align}
At energies far below the resonance, all $\tilde{U}$'s and $\tilde{m}^2$'s can be replaced by their bare values $U$ and $m^2$. 
At MiniBooNE or LSND only those terms contribute where the mass-squared difference is in the $\Delta m^2_{\mathrm{LSND}}$ region. 
These are the mass-squared differences ($\Delta m_{41}^2$, $\Delta m_{42}^2$, $\Delta m_{43}^2$), ($\Delta m_{51}^2$, $\Delta m_{52}^2$,
 $\Delta m_{53}^2$), and ($\Delta m_{61}^2$, $\Delta m_{62}^2$, $\Delta m_{63}^2$). The transition probability factorizes as:
\begin{align}
	P_{\nu_\mu\rightarrow\nu_e} \sim -4 \sin^2{\left(\Delta m_{\mathrm{LSND}}^2\frac{L}{2E}\right)}
	\left( \sum_{j=1,2,3} U_{\mu j}U_{e j} \right) \, \left( \sum_{j=4,5,6} U_{\mu j}U_{e j} \right) \, \text{.}
	\label{eq:prob6nu}
\end{align}
For simplicity, we define the mixing matrix as
\begin{align}
	U^{6\times 6} &= \underbrace{ ( U_{12}U_{13}U_{23} ) }_{U_0}  (  U_{14}U_{25}U_{36}) \nonumber \\
	&=U_0 \cdot  \begin{pmatrix}	c_\theta&0&0&s_\theta&0&0\\
		0&1&0&0&0&0\\
		0&0&1&0&0&0\\
		-s_\theta&0&0&c_\theta &0&0\\ 
		0&0&0&0&1&0\\ 
		0&0&0&0&0&1 \end{pmatrix}\cdot
	\begin{pmatrix}1&0&0&0&0&0\\ 
		0&c_\theta&0&0&s_\theta&0\\
		0&0&1&0&0&0\\
		0&0&0&1&0&0\\ 
		0& -s_\theta&0&0&c_\theta &0\\ 
		0&0&0&0&0&1 \end{pmatrix}\cdot
	\begin{pmatrix}1&0&0&0&0&0\\
		0&1&0&0&0&0\\
		0&0&c_\theta&0&0&s_\theta\\
		0&0&0&1&0&0\\0&0&0&0&1&0\\ 
		0&0&-s_\theta&0&0&c_\theta  \end{pmatrix}\nonumber\\
	&=U_0 \cdot \begin{pmatrix} 	c_\theta\cdot \mathbb{1}_{3\times3}& s_\theta \cdot \mathbb{1}_{3\times3}\\ 
		-s_\theta\cdot \mathbb{1}_{3\times3}& c_\theta \cdot \mathbb{1}_{3\times3}
	\end{pmatrix}\, \text{,}
	\label{mixingmatrix6n}
\end{align}
and we have the common mixing angle $\theta$ for $\theta_{14}=\theta_{25}=\theta$.

Since $U_0$ only describes a  rotation in the upper left corner, it can be written as
\begin{align}
	U_0=\begin{pmatrix} A_{3\times 3} & 0_{3\times 3}\\ 0_{3\times 3} & \mathbb{1}_{3\times 3} \end{pmatrix}\, \text{,}
	\label{eq:3times3}
\end{align}
Any submatrix formed from rotations alone is orthogonal, and therefore unitary.
So the sub-matrix $A_{3\times 3}$ is \textit{unitary}. This leads to the full mixing matrix
\begin{align}
	U^{6\times 6} = \begin{pmatrix} c_\theta \cdot A_{3\times 3} & s_\theta \cdot A_{3\times 3}\\ 
		-s_\theta \cdot \mathbb{1}_{3\times 3} & c_\theta \cdot\mathbb{1}_{3\times 3} 
	\end{pmatrix}\, \text{,}
	\label{eq:submatrix}
\end{align}
%
The oscillation probabilities for LSND and MiniBooNE in this model are generally 
\begin{align}
	P_{\nu_\mu\rightarrow\nu_e} \sim &-4 \sin^2{\left(\Delta m^2_{\mathrm{LSND}}\frac{L}{2E}\right)} \times \nonumber\\
	&\times \left( U_{\mu 1}U_{e 1}+U_{\mu 2}U_{e 2}+U_{\mu 3}U_{e 3}\right)
	\left(U_{\mu 4}U_{e 4}+U_{\mu 5}U_{e 5}+U_{\mu 6}U_{e 6}\right) \, \text{.}
	\label{eq:democraticCancelation}
\end{align}
Since the submatrix $A_{3\times 3}$ itself is unitary and the unitarity conditions $\sum_k^6 U_{\mu k}U_{e k} = 0$ as well as
$\sum_k^3 A_{\mu k}A_{e k} = 0$ hold, it is readily seen that both brackets have to vanish when all 
new mixing angles $\theta_{ij}$ are the same.
(The first bracketed term is $\cos^2\theta \, (A_{3\times 3}^{(\mu{\rm\  row})} \cdot A_{3\times 3}^{(e{\rm\ row})}=0 )$ and 
the second bracketed term is $\sin^2\theta \, (A_{3\times 3}^{(\mu{\rm\ row})}  \cdot A_{3\times 3}^{(e{\rm\ row})}=0 )$.
Consequently, a $3+3\nu$ model with three additional sterile neutrinos and a common 
resonance energy also fails.
On the one hand it is indeed possible to avoid the constraints from atmospheric neutrinos above the resonance, if all three 
sterile neutrinos mix with the same mixing angle. This removes
the additional mass-squared differences in the considered region and thereby immunizes the model against 
constraints from high energy atmospheric neutrinos. 
On the other hand, however, the democratic mixing simultaneously implies a 
vanishing transition amplitude for $\overset{\scriptscriptstyle(-)}{\nu}_\mu\rightarrow\overset{\scriptscriptstyle(-)}{\nu}_e$ oscillations at MiniBooNE 
or LSND and consequently invalidates the desired main feature of the model. 
As we next show, the issue can be resolved by assigning different resonance energies to the different 
sterile neutrinos. 

\subsection{Behavior Below the Resonance}\label{sec:PhenoBelowResonance}
Although the transition 
$\overset{\scriptscriptstyle(-)}{\nu}_\mu\rightarrow\overset{\scriptscriptstyle(-)}{\nu}_{e/\tau}$ vanishes far below the resonance,
$\overset{\scriptscriptstyle(-)}{\nu}_\mu\to\overset{\scriptscriptstyle(-)}{\nu}_{s1,2,3}$ 
does not vanish due to the bare mixing. The same is also true for 
$\overset{\scriptscriptstyle(-)}{\nu}_e\to\overset{\scriptscriptstyle(-)}{\nu}_{s1,2,3}$ and
$\overset{\scriptscriptstyle(-)}{\nu}_\tau\to\overset{\scriptscriptstyle(-)}{\nu}_{s1,2,3}$.
This is particularly interesting for reactor experiments, which usually operate in the MeV-region, since they are predicted by this model to measure a deviation 
in the $\bar{\nu}_e\to\bar{\nu}_e$ channel. 
A good approximation for 
$\overset{\scriptscriptstyle(-)}{\nu}_e\rightarrow\overset{\scriptscriptstyle(-)}{\nu}_e$
in the low energy region well
below the resonances is given by (compare Eq.~\eqref{eq:democraticCancelation}):
\begin{align}
	P_{\nu_e\rightarrow\nu_e} \sim &1-4 \sin^2{\left(\Delta m^2_{\mathrm{LSND}}\frac{L}{2E}\right)} 
	\left( U_{e1}^2+U_{e 2}^2+U_{e 3}^2\right)
	\left(U_{e 4}^2+U_{e 5}^2+U_{e 6}^2\right)\nonumber\\
	=& 1-4 \sin^2{\left(\Delta m^2_{\mathrm{LSND}}\frac{L}{2E}\right)}\cos^2{\theta}\sin^2{\theta}\nonumber\\
	=& 1-\sin^2{\left(\Delta m^2_{\mathrm{LSND}}\frac{L}{2E}\right)}\sin^2{2\theta} \, \text{,}
	\label{}
\end{align}
where again the unitarity conditions are used. This expression resembles a simple $3+1\nu$ model for disappearance experiments in
the low energy region, which is actually favored by the Reactor- or Gallium anomalies.

\subsection{Different Effective Potentials for Different Sterile Neutrinos}
From the previous discussion it becomes clear that both the low energy limit $E\to 0$ and the high energy limit $E\to \infty$ 
are independent of the specific values of the effective potentials of the different sterile neutrinos. What matters is only 
that the energy is well below or well above the respective resonance energy. 
However, there exists the possibility of assigning different resonance energies to the sterile neutrinos
(e.g.~by tying each sterile neutrino to its own extra dimension). 
The potentials for the neutrinos do not necessarily have to be the same for each sterile neutrino. 
If the effective potentials differ, we still expect 
a resonant behavior around the resonance energy also for
$\overset{\scriptscriptstyle(-)}{\nu}_\mu\rightarrow\overset{\scriptscriptstyle(-)}{\nu}_e$
transitions:
In the intermediate energy region the arguments made in the previous chapter no longer hold since the effective mixing 
angles differ for the different sterile neutrinos as a consequence of to the different effective potentials. 
As long as the bare mixing angle is
the same for all sterile neutrinos, we nevertheless end up with the aforementioned low and high energy behavior.
The effective potential \footnote{$V_\text{eff}$ denotes the potential due to the ADR itself. This potential is ubiquitous and does not distinguish between neutrinos and anti-neutrinos, since it is an effect of spacetime itself. In the presence of matter, additional terms from coherent forward scattering have to be added to $V_\text{eff}$.} with different ADR parameters can be written as
\begin{align}
	V_{\mathrm{eff}}= \begin{pmatrix} 	0&0&0&0&0&0\\ 
		0&0&0&0&0&0\\ 
		0&0&0&0&0&0\\ 
		0&0&0&\eta \cdot \varepsilon E&0&0\\
		0&0&0&0&\kappa \cdot \varepsilon E &0\\
		0&0&0&0&0&\xi \cdot \varepsilon E\end{pmatrix}\text{.}
	\label{}
\end{align}
Note that we still only find three independent new parameters in the potential, since an overall factor shared by $\kappa, \, \xi$ and $\eta$ can be absorbed into $\epsilon$, so that one of the parameters can always be chosen as $1$ without loss of generality.

It has become clear from the previous discussion that a model with a different ADR for each additional neutrino and a common mixing angle $\theta$ potentially offers a possibility to resolve some of the long standing neutrino anomalies without being ruled out by long baseline experiments. 

\section{Phenomenology}\label{sec:Pheno}
In this section we study several benchmark points (BMPs) and show the phenomenological features of the model in each of these regimes. We will briefly discuss the structure of the $\Delta m^2$ as a function of the neutrino energy for each BMP. Furthermore, we will focus mainly on the short baseline appearance experiments MiniBooNE and LSND, low energy disappearance experiments NEOS and KamLAND, and long baseline experiment T2K. We will, however, also refer to KARMEN as well as atmospheric neutrino experiments e.g. IceCube and SuperKamiokande.

In the numerical analysis we adopt the best-fit values from \cite{Olive:2016xmw} for the standard $3\nu$ mixing angles and mass-squared differences,
and assume normal ordering and vanishing CP violation). We also neglect matter effects for every short baseline experiment, since they do not solve the problem we intend to address. The best-fit values are
\begin{align}
	\Delta m_{21}^2 &= 7.37\times 10^{-5}\text{eV}^2\, \text{,} 	\quad
	\Delta m_{31}^2 = 2.50\times 10^{-3}\text{eV}^2\, \text{,}  \nonumber \\
	\sin^2{\theta_{12}}&=0.297\, \text{,} \quad 
	\sin^2{\theta_{13}}=0.0214\,\, \text{,} \quad
	\sin^2{\theta_{23}}=0.437 \text{.} 
	\label{eq:3nuMixing}
\end{align}

Interpreting the MiniBooNE result as a resonance excess, there are plenty of possibilities to realize a resonance at the MiniBooNE scale, since $E_\text{Res} \sim \sqrt{\frac{\cos{2\theta} \Delta m^2}{\varepsilon} }$ depends on three BSM parameters. In this work, we show a selection of BMPs from the $\Delta m^2 = \mathcal{O}(1\, \text{eV}^2)$ and $\Delta m^2 = \mathcal{O}(10\, \text{eV}^2)$ regimes.

We will explicitly show BMPs, where some of the current anomalies can be resolved and others cannot. This section aims to provide the reader with an intuition, which implications on the broader picture the parameters present in this model have. Also, in the case of an anomaly turning out to be a statistical fluctuation, an underestimation of a systematic error or not well enough understood nuclear physics effects, this model can still provide a good explanation for the global oscillation picture.

\begin{table}
	\centering
	\begin{tabular}{c|c|c|c|c}
		BMP  & MiniBooNE & LSND & Ga/Reactor & Consistency with constraints \\ \hline 
		BMP1 & \cmark & \cmark & \xmark & \xmark \\
		BMP2 & \xmark & \xmark & \cmark & \cmark \\
		BMP3 & \cmark & \xmark & \xmark & \cmark \\
		BMP4 & \cmark & \cmark & \textcolor{orange}{?} & \cmark \\
	\end{tabular}
	\caption{Overview of benchmark points (BMPs) discussed in this work. The corresponding parameter sets illustrate the characteristic features of the scenario and point towards a strategy to obtain fits to various neutrino anomalies.}
	\label{tab:BMPs}
\end{table}

The phenomenology of the various BMPs is summarized in table \ref{tab:BMPs}. While these four BMPs are far from exhausting the five-dimensional parameter space of the scenario presented, they illustrate characteristic features and point out strategies to obtain a good fit to at least some of the neutrino anomalies. A more elaborate data analysis would require access to proprietary data of the experimental collaborations and is beyond the scope of this work. 

\subsection{The $\Delta m^2 = \mathcal{O}(1\, \text{eV}^2)$ Regime}\label{sec:1eV2}
Driven by the '$1\,\text{eV}^2$-miracle', which arose due to the LSND, MiniBooNE and Reactor/Gallium-results, we first explore the $\Delta m^2 = \mathcal{O}(1\, \text{eV}^2)$ regime in this model. We present two BMPs and discuss their viability.
\subsubsection{Benchmark Point 1}
The BSM parameters of this BMP are 
\begin{align}
	\begin{split}
		\Delta m_\text{LSND}^2 &= 1.5\, \text{eV}^2 \, ,\sin^2{\theta}=10^{-2}\, ,  \\ 
		\varepsilon= 8 \cdot 10^{-18} \, , \eta &= 1  \, , \kappa= 150 \, , \xi= 150 \\
		\Rightarrow E_{\text{Res},\eta}&=303.1 \, \text{MeV} \, , \\
		E_{\text{Res},\kappa}&= 24.7 \, \text{MeV} \, , \\
		E_{\text{Res},\xi}&=24.7 \, \text{MeV} \,  . 
	\end{split}
	\label{eq:BMP1}
\end{align}

\begin{figure}
	\centering
	\subfigure[$P_{\nu_\mu\rightarrow\nu_e}$]{\includegraphics[width=0.44\textwidth]{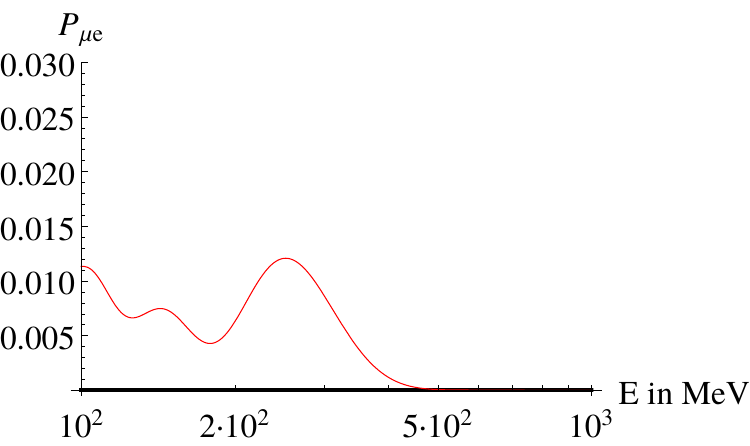}}
	\subfigure[$P_{\nu_\mu\rightarrow\nu_\mu}$]{\includegraphics[width=0.55\textwidth]{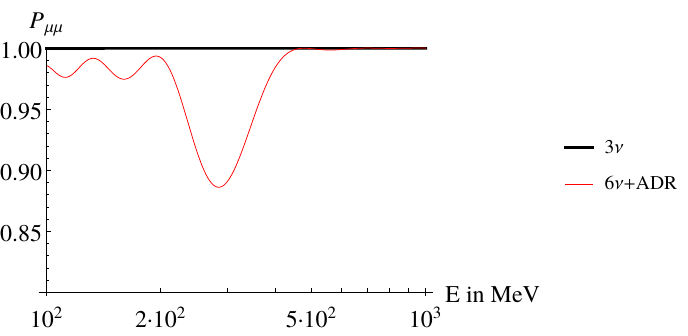}}
	\caption{Appearance (a) and disappearance (b) probabilities at MiniBooNE for BMP 1.}
	\label{fig:MiniBooNEBMP1}
\end{figure}

\begin{figure}
	\centering
	\subfigure[$P_{\nu_\mu\rightarrow\nu_e} @ \text{LSND}$]{\includegraphics[width=0.44\textwidth]{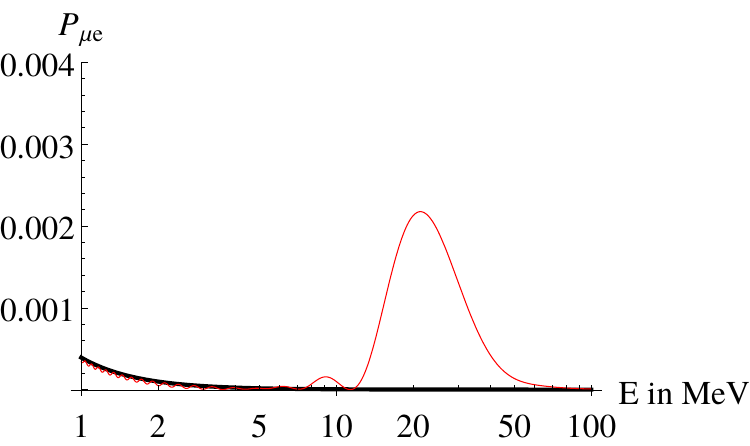}}
	\subfigure[$P_{\nu_\mu\rightarrow\nu_e} @ \text{KARMEN}$]{\includegraphics[width=0.55\textwidth]{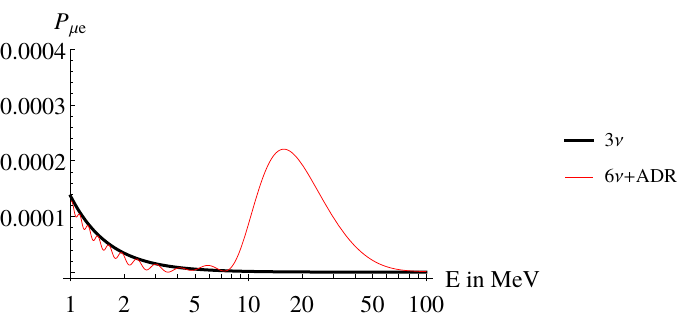}}
	\caption{LSND (a) and KARMEN (b) oscillation probabilities for BMP 1.}
	\label{fig:LSNDKARMENBMP1}
\end{figure}
As one can see in figs. \ref{fig:MiniBooNEBMP1} and \ref{fig:LSNDKARMENBMP1} we plot the oscillation probability at the short baseline experiments MiniBooNE, LSND and KARMEN as a function of energy. LSND features a resonant behavior, while we find the same characteristics at KARMEN but about one order of magnitude less prominent. This is a desired behavior, since LSND was able to find an excess, whereas KARMEN could not confirm the discovery. At MiniBooNE we can observe a broad resonance peak at $~250\,$MeV, while the suppression of appearance probability due to the decoupling of the sterile states can be observed at $\sim 400\,$MeV. Since MiniBooNE is blind to lower values than $200\,$MeV, the experiment is unable to determine the broadness of the resonance extending to lower energies. 

\begin{figure}
	\centering
	\subfigure[$P_{\nu_e \rightarrow\nu_e} @ \text{NEOS}$]{\includegraphics[width=0.49\textwidth]{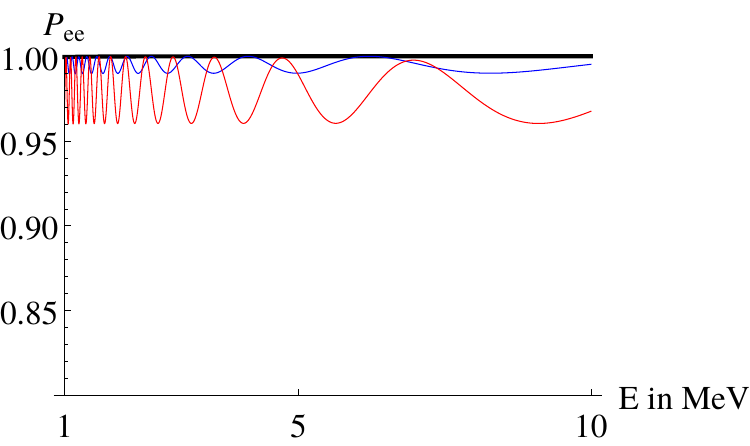}}
	\subfigure[$P_{\nu_e \rightarrow\nu_e} @ \text{KamLAND}$]{\includegraphics[width=0.49\textwidth]{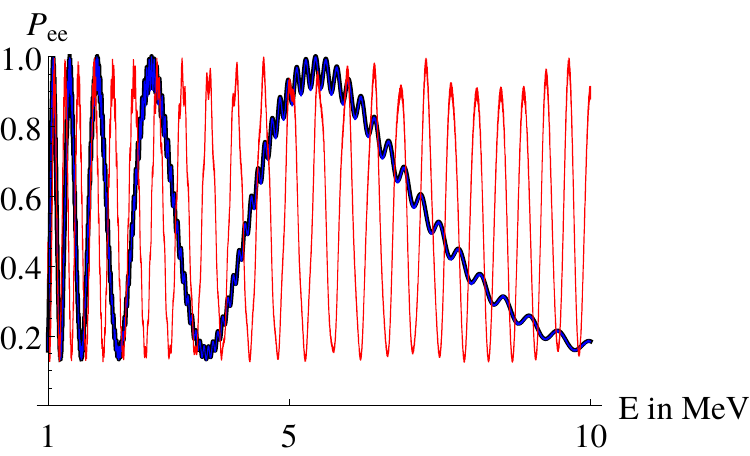}} \\
	\subfigure[$P_{\nu_e \rightarrow\nu_e} @ \text{Daya-Bay}$]{\includegraphics[width=0.63\textwidth]{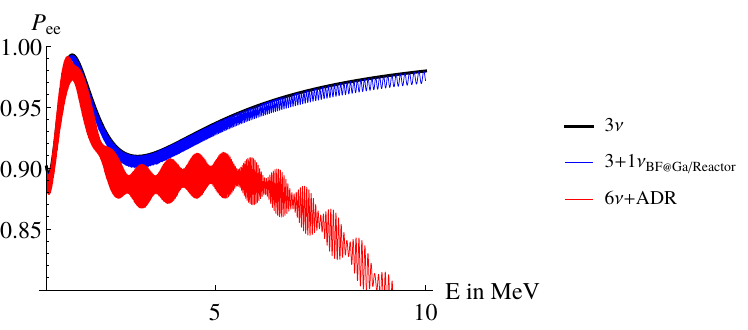}}
	\caption{Disappearance probabilities at NEOS (a), KamLAND (b) and Daya-Bay (c) for BMP 1.
	The blue curve shows the global best fit point for a 3+1-scenario for reactor and gallium experiments found by \cite{Dentler:2018sju} with $\Delta m^2_\text{BF}=1.3\,\text{eV}^2$ and $\sin^2{2\theta_{14}}=0.01$.}
	\label{fig:ReactorsBMP1}
\end{figure}

This particular BMP, however, while possibly providing a good fit to both LSND and MiniBooNE, is already excluded by other experiments, since it does not satisfy the bounds posed by reactor experiments such as KamLAND (see fig. \ref{fig:ReactorsBMP1}). 

\begin{figure}
	\centering
	\includegraphics[width=0.8\textwidth]{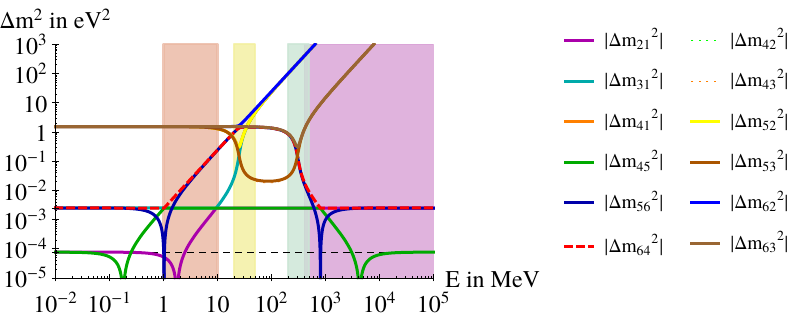}
	\caption{Effective $\Delta m^2$ depending on the energy $E$ for BMP 1. The red region corresponds to the energy range of reactor and gallium experiments, the yellow region represents LSND, while the green and purple regions represent MiniBooNE and long baseline experiments respectively.}
	\label{fig:dmSquaredBMP1}
\end{figure}

The whole scenario can be understood in more detail by looking at the corresponding $\Delta m^2(E)$, which can be seen in fig. \ref{fig:dmSquaredBMP1}. In this plot, we show all possible combinations of differences of squared mass eigenvalues $|\Delta m^2_{ij}|$ that appear in the system over several orders of magnitude. We indicate the energy ranges of the experiments that are operating at the moment with colored regions. Additionally, we include the standard 3$\nu$-mass gaps $\Delta m^2_{21}$ and $\Delta m^2_{31}$ as points of reference.  

While the equations developed in the sec. \ref{sec:PhenoBelowResonance} still hold at low energies, the effective mass gaps have not converged at the reactor scale ($\mathcal{O}(1\, \text{MeV})$) yet. 
We can see in fig. \ref{fig:dmSquaredBMP1} that the specific ordering of the resonance energies leads to a peculiar kind of level crossing, where certain eigenvalues of the hamiltonian obtain the same numerical values and therefore the $\Delta m^2$s drop to zero.

\begin{figure}
	\centering
	\includegraphics[width=0.8\textwidth]{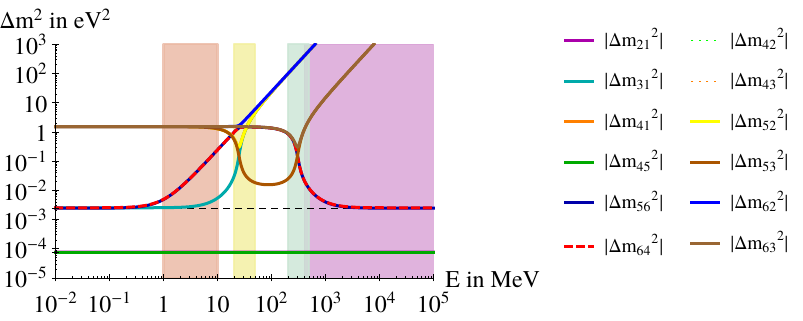}
	\caption{Effective $\Delta m^2$ depending on the energy $E$ for BMP 1 with re-ordered ADR parameters $\eta > \kappa > \xi$.}
	\label{fig:dmSquaredBMP1V2}
\end{figure}

This can be avoided by re-ordering the ADR parameters to $\eta > \kappa > \xi$. In this case, only non-vanishing $\Delta m^2$s are present, which can help improve the convergence to the standard mass gaps $\Delta m^2_{21}$ and $\Delta m^2_{31}$ in the transition regimes to high/low energies significantly. We can see in fig. \ref{fig:dmSquaredBMP1V2} that convergence is reached faster than 
in fig. \ref{fig:dmSquaredBMP1}. It is also visible, however, that this BMP is still not viable, since in the long baseline as well as the reactor regime the convergence of the $\Delta m^2$s to the 3+1$\nu$/3$\nu$ pictures does not happen fast enough.
Another constraint arises from a potential distortion of the solar neutrino spectrum. In particular the spectrum of ${}^8$B neutrinos has its maximum at a comparably large energy of about $8\,$MeV and an end point energy of about $14\,$MeV.
A possibility of taking measures to increase convergence is lowering the mixing angle, since in this case the level crossings occur in a steeper slope.

 \FloatBarrier

\subsubsection{Benchmark Point 2} 
This BMP features BSM parameters with re-ordered ADR parameters as
\begin{align}
	\begin{split}
		\Delta m_\text{LSND}^2 &= 1.3\, \text{eV}^2 \, ,\sin^2{\theta}=2\cdot 10^{-3}\, ,  \\ 
		\varepsilon= 1.5 \cdot 10^{-17} \, , \eta &= 10.9 \, , \kappa= 10.7 \, , \xi= 1 \\
		\Rightarrow E_{\text{Res},\eta}&=62.9 \, \text{MeV} \, , \\
		E_{\text{Res},\kappa}&= 63.5 \, \text{MeV} \, , \\
		E_{\text{Res},\xi}&=207.8 \, \text{MeV} \,  .
	\end{split}
	\label{eq:BMP2}
\end{align}

\begin{figure}
	\centering
	\includegraphics[width=0.8\textwidth]{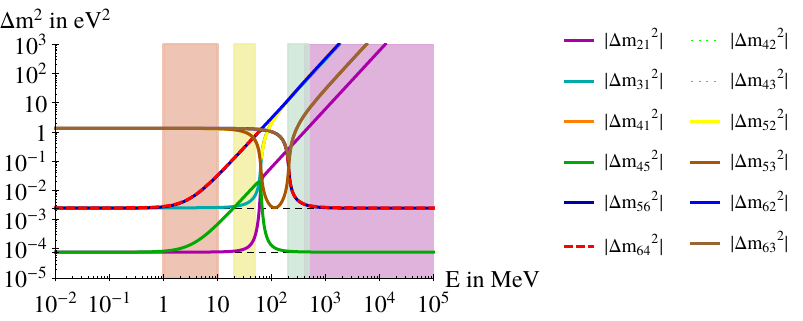}
	\caption{Effective $\Delta m^2$ depending on the energy $E$ for BMP 2.}
	\label{fig:dmSquaredBMP2}
\end{figure}

The corresponding $\Delta m^2(E)$ can be seen in fig. \ref{fig:dmSquaredBMP2}. 
Featuring a mixing, which is an order of magnitude lower than in BMP1, this BMP has the potential to meet all bounds from reactor and long baseline experiments. Fig. \ref{fig:dmSquaredBMP2} shows that in the reactor regime there is no interfering $\Delta m^2$ in the region of $\Delta m^2_{21}$ so that the oscillation probabilities in low energy disappearance experiments retain their shape up to a typical 3+1$\nu$-like modulation, which is favorable in the $\mathcal{O}(1 \, \text{eV}^2)$ case.  

\begin{figure}
	\centering
	\subfigure[$P_{\nu_\mu\rightarrow\nu_e}$]{\includegraphics[width=0.44\textwidth]{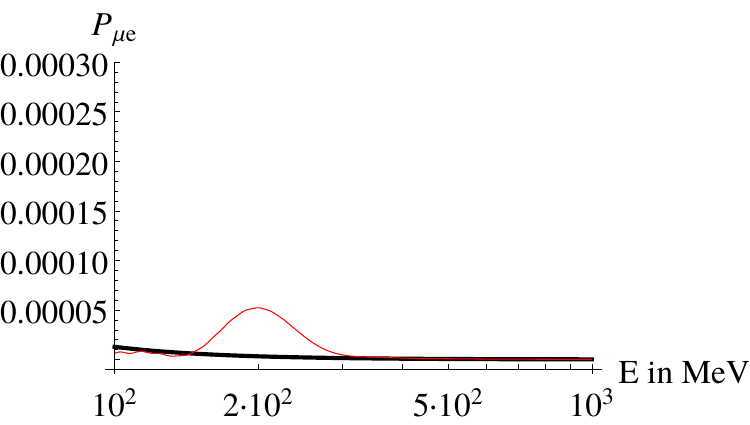}}
	\subfigure[$P_{\nu_\mu\rightarrow\nu_\mu}$]{\includegraphics[width=0.55\textwidth]{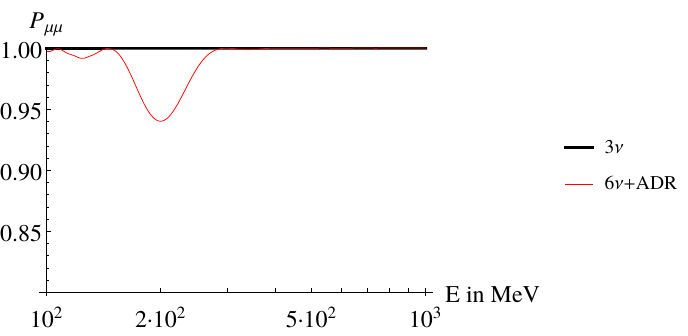}}
	\caption{Appearance (a) and disappearance (b) probabilities at MiniBooNE for BMP 2.}
	\label{fig:MiniBooNEBMP2}
\end{figure}

\begin{figure}[H]
	\centering
	\subfigure[$P_{\nu_\mu\rightarrow\nu_e} @ \text{LSND}$]{\includegraphics[width=0.44\textwidth]{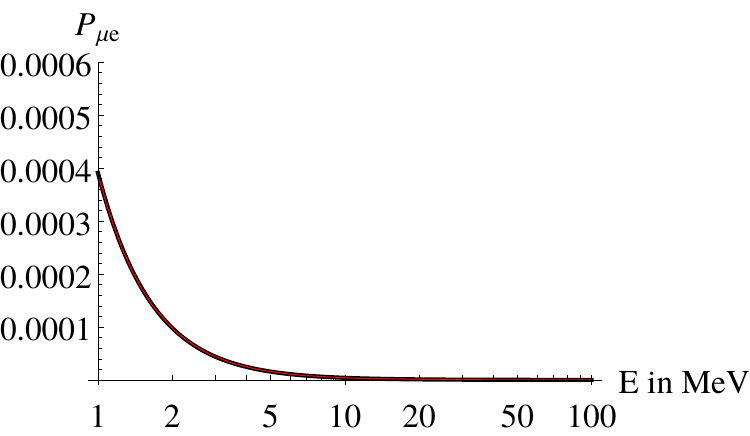}}
	\subfigure[$P_{\nu_\mu\rightarrow\nu_e} @ \text{KARMEN}$]{\includegraphics[width=0.55\textwidth]{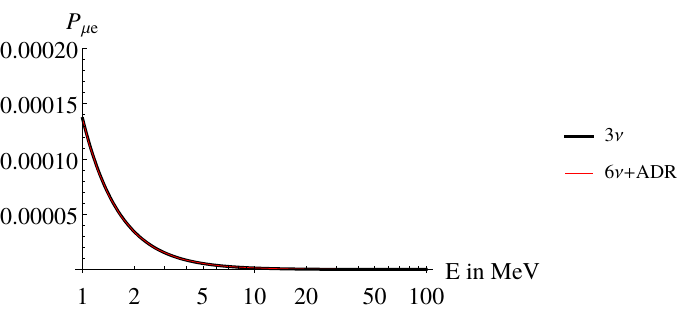}}
	\caption{LSND (a) and KARMEN (b) oscillation probabilities for BMP 2.}
	\label{fig:LSNDKARMENBMP2}
\end{figure}

The oscillation probability at MiniBooNE (fig. \ref{fig:MiniBooNEBMP2}), although there is a level-crossing at $\sim 250\,$MeV, almost vanishes, since the conversion rate from sterile to active neutrinos is suppressed due to the lower mixing compared to BMP1. LSND (fig. \ref{fig:LSNDKARMENBMP2}) suffers from the same vanishing behavior, since the mechanism for appearance is the same there. This means that this benchmark cannot explain the SBL anomalies, but still is a candidate to agree the reactor results with the long baseline experiments, should the SBL anomalies prove to be of non-oscillatory origin or a statistical fluctuation. These prerequisites make it worthwhile to still take a closer look at the probabilities at reactors as well as LBL and atmospheric experiments. 

\begin{figure}
	\centering
	\subfigure[$P_{\nu_e \rightarrow\nu_e} @ \text{NEOS}$]{\includegraphics[width=0.49\textwidth]{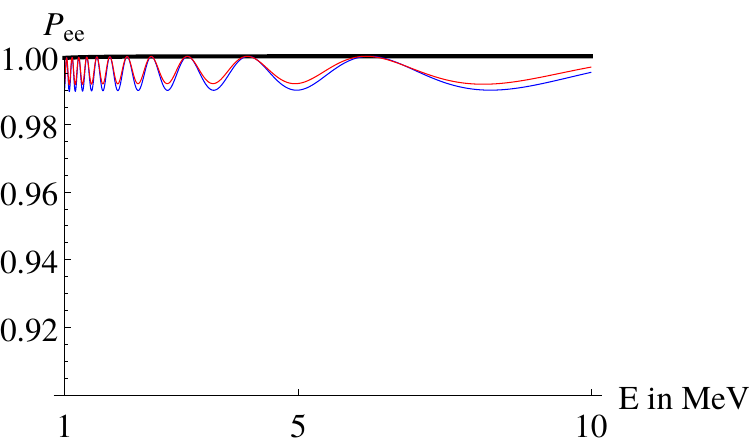}}
	\subfigure[$P_{\nu_e \rightarrow\nu_e} @ \text{KamLAND}$]{\includegraphics[width=0.49\textwidth]{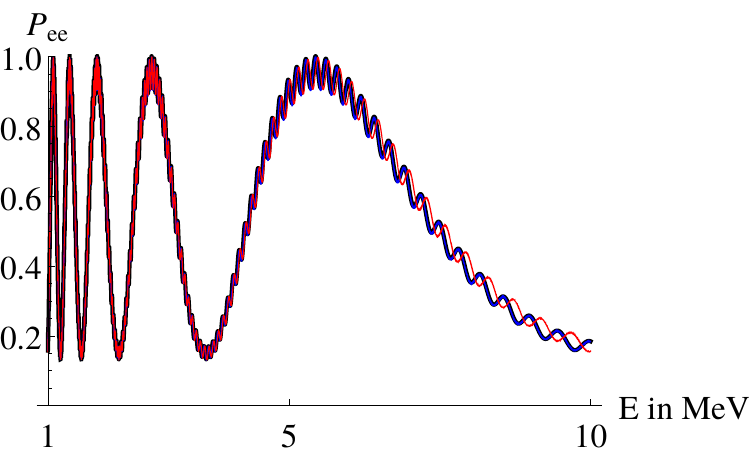}} \\
	\subfigure[$P_{\nu_e \rightarrow\nu_e} @ \text{Daya-Bay}$]{\includegraphics[width=0.63\textwidth]{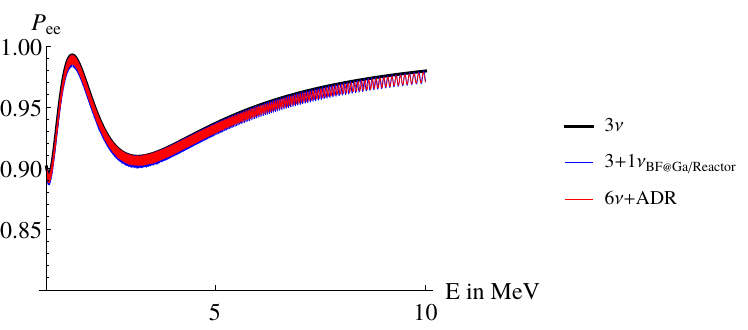}}
	\caption{Disappearance probabilities at NEOS (a), KamLAND (b) and Daya-Bay (c) for BMP 2.
	The blue curve shows the global best fit point for a 3+1-scenario for reactor and gallium experiments found by \cite{Dentler:2018sju} with $\Delta m^2_\text{BF}=1.3\,\text{eV}^2$ and $\sin^2{2\theta_{14}}=0.01$.}
	\label{fig:ReactorsBMP2}
\end{figure}

Since the same mass gap of $\Delta m^2_\text{LSND}=1.3\,\text{eV}^2$ is employed, the global BF scenario found in 
\cite{Dentler:2018sju} and BMP2 fit very well at reactor experiments, which can be seen in fig. \ref{fig:ReactorsBMP2}. This doesn't come as a surprise, given that convergence is fairly well achieved at this point already and the equations of sec. \ref{sec:PhenoBelowResonance} govern this energy range. 

\begin{figure}[H]
	\centering
	\subfigure[$P_{\nu_\mu \rightarrow\nu_\mu}$]{\includegraphics[width=0.44\textwidth]{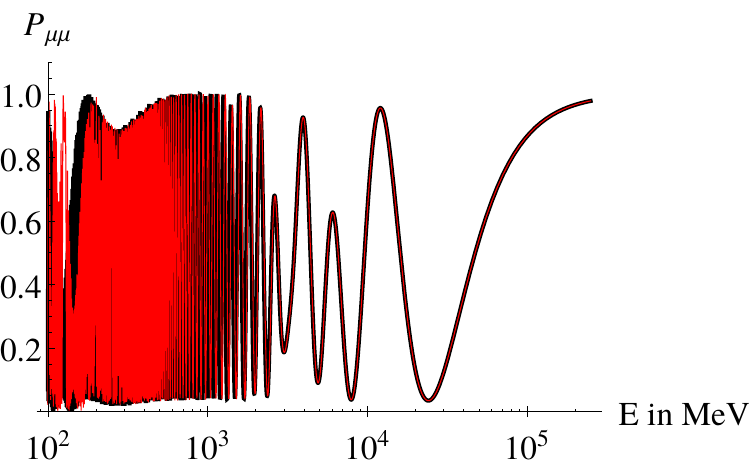}}
	\subfigure[$P_{\nu_\mu \rightarrow\nu_e}$]{\includegraphics[width=0.55\textwidth]{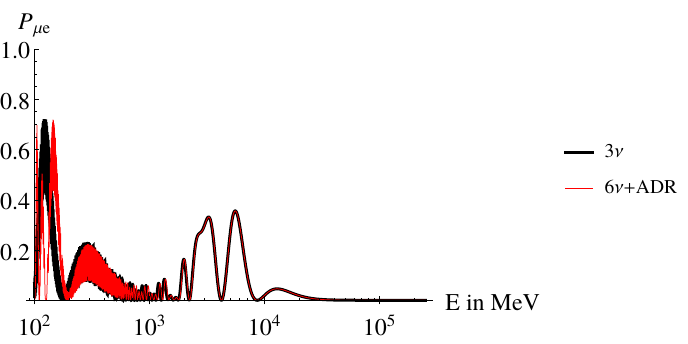}} 
	\caption{Disappearance and appearance probabilities of upward going neutrinos at atmospheric experiments for BMP 2 , where $L \sim d_\text{Earth}$.}
	\label{fig:AtmosphericBMP2}
\end{figure}

Since most models that include light sterile neutrinos have a serious problem with oscillations at long baseline experiments, we take a closer look at oscillation probability at long baseline experiments like T2K or the atmospheric neutrino experiments. We show disappearance as well as appearance probabilities in fig. \ref{fig:AtmosphericBMP2}. For the upward-going neutrinos we incorporated a mantle-core-mantle (MCM) model for the density profile of the Earth, where the density of the mantle is $\rho_M \approx 4.66 \frac{\text{g}}{\text{cm}^3}$ and the density of the core is $\rho_C \approx 11.8 \frac{\text{g}}{\text{cm}^3}$. The MCM model is a reasonably good approximation to the best model of the Earth's density that is available, the Preliminary Reference Earth Model (PREM) \cite{DZIEWONSKI1981297}, averaging the matter density over the mantle and core. 

\begin{figure}
	\centering
	\subfigure[$P_{\nu_\mu \rightarrow\nu_\mu}$]{\includegraphics[width=0.44\textwidth]{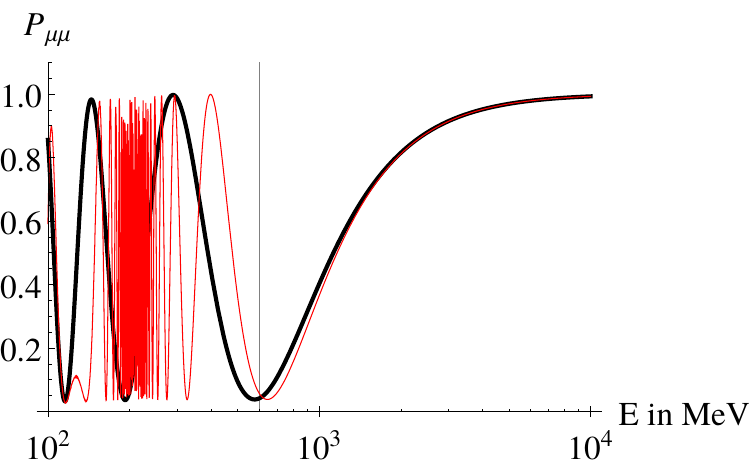}}
	\subfigure[$P_{\nu_\mu \rightarrow\nu_e}$]{\includegraphics[width=0.55\textwidth]{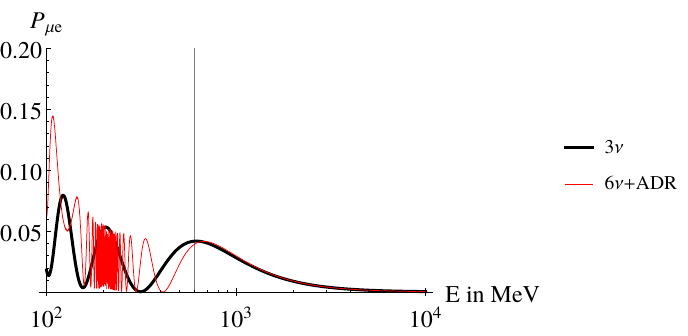}} \\
	\subfigure[$P_{\nu_\mu \rightarrow\nu_\mu}$]{\includegraphics[width=0.44\textwidth]{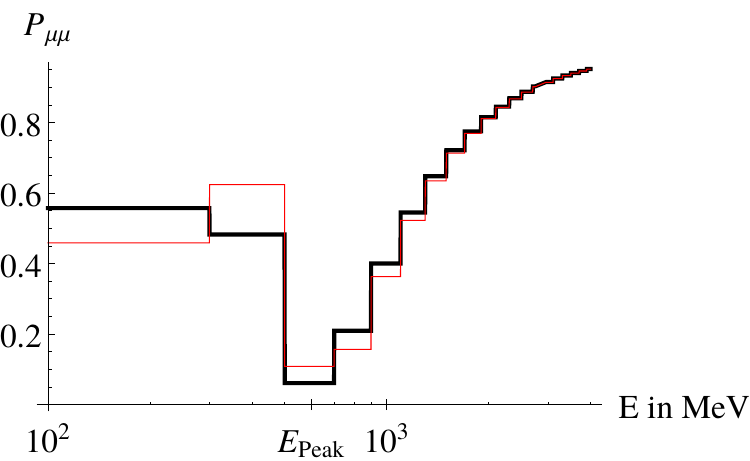}}
	\subfigure[$P_{\nu_\mu \rightarrow\nu_e}$]{\includegraphics[width=0.55\textwidth]{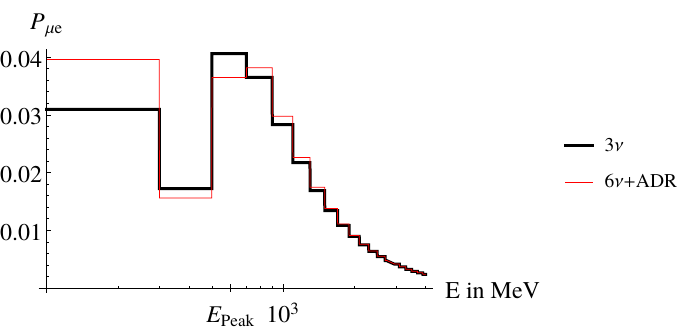}}
	\caption{Disappearance a) and appearance b) probabilities at T2K for BMP 2 and their corresponding binned versions c) and d) with a bin-size of $200\,$MeV. The grey line indicates the peak energy of the accelerators neutrino spectrum.}
	\label{fig:T2KBMP2}
\end{figure}

We find that the upward going neutrino oscillation probability, both in appearance and disappearance is in good agreement with the standard 3$\nu$ theory. This is a trait, which vanilla-type 3+1$\nu$ models fail to achieve.
Since T2K is the accelerator experiment with the lowest peak energy ($\sim 600\,$MeV) of it's neutrino energy spectrum, we consider it as the long baseline accelerator experiment with the most potential to constrain ADR models of this kind. It is most sensitive to the convergence gradient, which seems to be a crucial point in the models effectiveness to successfully explain non-observation of high energy active-sterile oscillation. 
In the T2K plots in fig. \ref{fig:T2KBMP2}, matter effects of the Earth's crust are included, where the matter density is approximated as constant over the course of the whole baseline. 

On the oscillation probability level, the reader might get the impression that the chosen BMP is excluded by T2K, but the limited energy resolution of T2K in combination with avergaging effects over fast oscillations has to be taken into account at this point. Figs. \ref{fig:T2KBMP2} c) and d) show the corresponding bin-plots with a bin-size of $200\,$MeV, where it can be seen that the deviation from the standard theory is small. As an accelerator experiment whose beam has a distinct energy distribution with a distinct peak at $\sim 600\,$MeV, there is little statistics at energies below the peak energy. We argue that a dedicated fit of this model to the T2K data is neccessary to be able to 
constrain the model, since averaging effects can open windows within the parameter space, where a good fit on probability level is possible.
As the value of $\Delta m^2_{21}$ isn’t affected for energies below $20-30\,$MeV, for BMP 2 no significant distortion of solar neutrino spectra is expected.

\subsection{The $\Delta m^2 = \mathcal{O}(10\, \text{eV}^2)$ Regime}\label{sec:10eV2}
In this section, we study the regime where $\Delta m^2 = \mathcal{O}(10\, \text{eV}^2)$. We present two BMPs, which provide a resonance at MiniBooNE, while simultaneously not violating the bounds of long baseline and low energy disappearance experiments. To keep the resonance energy for MiniBooNE at $\sim 250\,$MeV, the mixing angles are chosen lower than the ones in sec. \ref{sec:1eV2}.

 \FloatBarrier

\subsubsection{Benchmark Point 3}
The BSM parameters of this BMP are 

\begin{align}
	\begin{split}
		\Delta m_\text{LSND}^2 &= 30\, \text{eV}^2 \, ,\sin^2{\theta}=10^{-4}\, ,  \\ 
		\varepsilon= 3 \cdot 10^{-16} \, , \eta &= 10.1 \, , \kappa= 10 \, , \xi= 1 \,\\
		\Rightarrow E_{\text{Res},\eta}&=70.3\, \text{MeV} \, , \\
		E_{\text{Res},\kappa}&= 70.7\, \text{MeV} \, , \\
		E_{\text{Res},\xi}&=223.6\, \text{MeV} \,  .
	\end{split}
	\label{eq:BMP3}
\end{align}

\begin{figure}
	\centering
	\includegraphics[width=0.8\textwidth]{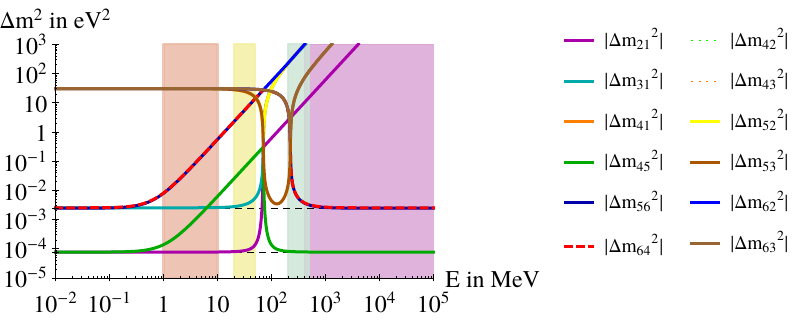}
	\caption{Effective $\Delta m^2$ depending on the energy $E$ for BMP 3.}
	\label{fig:dmSquaredBMP3}
\end{figure}

\begin{figure}[H]
	\centering
	\subfigure[$P_{\nu_\mu\rightarrow\nu_e}$]{\includegraphics[width=0.44\textwidth]{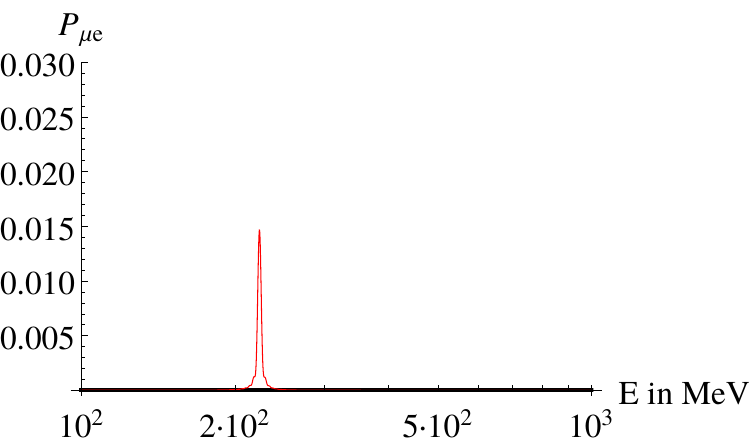}}
	\subfigure[$P_{\nu_\mu\rightarrow\nu_\mu}$]{\includegraphics[width=0.55\textwidth]{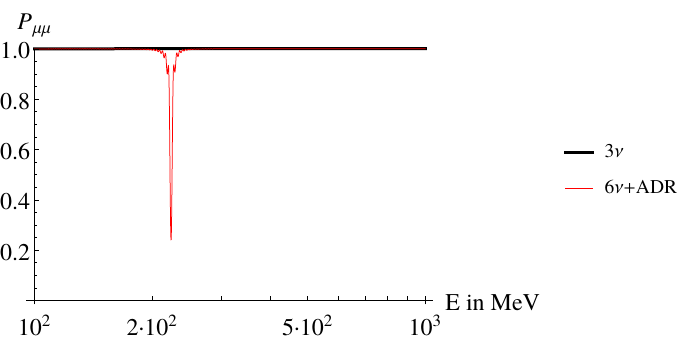}}
	\caption{Appearance (a) and disappearance (b) probabilities at MiniBooNE for BMP 3.}
	\label{fig:MiniBooNEBMP3}
\end{figure}

\begin{figure}
	\centering
	\subfigure[$P_{\nu_\mu\rightarrow\nu_e} @ \text{LSND}$]{\includegraphics[width=0.44\textwidth]{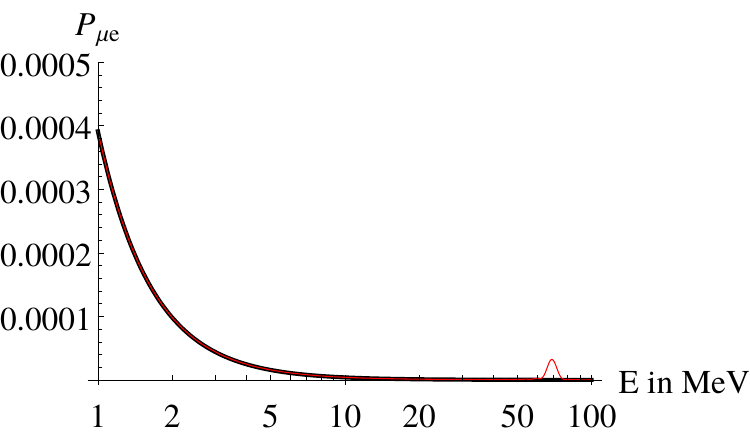}}
	\subfigure[$P_{\nu_\mu\rightarrow\nu_e} @ \text{KARMEN}$]{\includegraphics[width=0.55\textwidth]{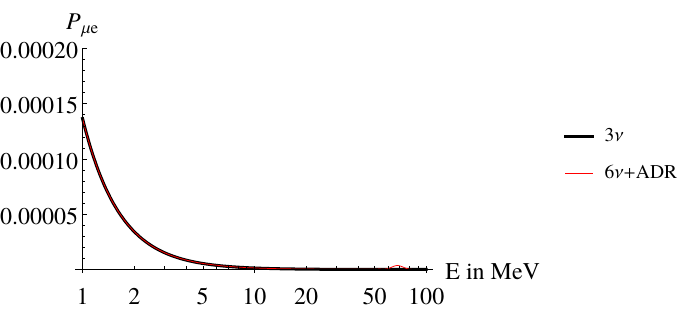}}
	\caption{LSND (a) and KARMEN (b) oscillation probabilities for BMP 3.}
	\label{fig:LSNDKARMENBMP3}
\end{figure}

The corresponding $\Delta m^2(E)$ can be seen in fig. \ref{fig:dmSquaredBMP3}. 
Since the mixing angles are very low in this scenario, the level crossings happen considerably fast leading to a steep slope in the $\Delta m^2 \text{vs.} E$-diagram. This also naturally leads to very distinct resonance peaks at these level crossings. Whether the MiniBooNE experiment actually favors a narrow or a more broad type of resonance must be determined by a dedicated fit of this model to the data. 
The presented BMP aims not to solve the LSND anomaly (see fig. \ref{fig:LSNDKARMENBMP3}), as the resonance energy is not in the correct ballpark for that. Fig. \ref{fig:MiniBooNEBMP3} reveals, though, that MiniBooNE does show a distinct resonance. 

\begin{figure}
	\centering
	\subfigure[$P_{\nu_e \rightarrow\nu_e} @ \text{NEOS}$]{\includegraphics[width=0.49\textwidth]{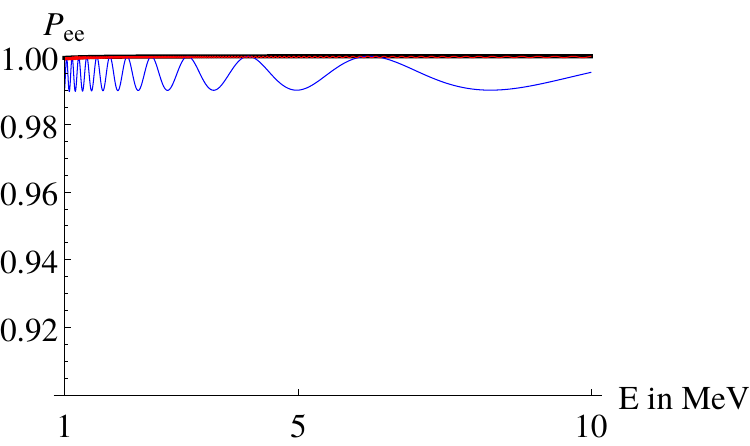}}
	\subfigure[$P_{\nu_e \rightarrow\nu_e} @ \text{KamLAND}$]{\includegraphics[width=0.49\textwidth]{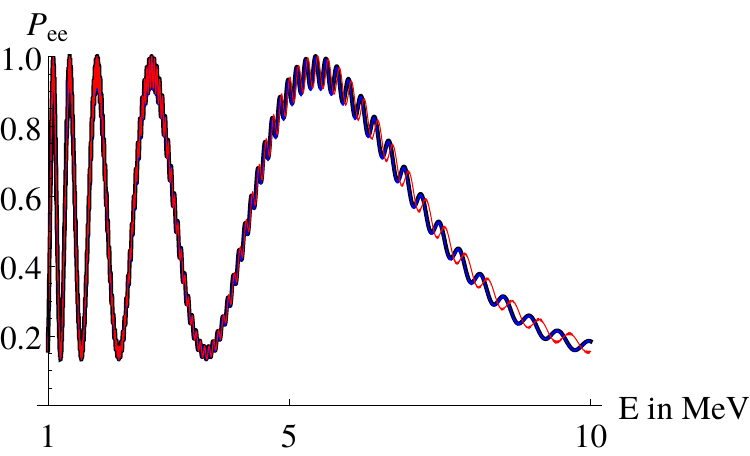}} \\
	\subfigure[$P_{\nu_e \rightarrow\nu_e} @ \text{Daya-Bay}$]{\includegraphics[width=0.63\textwidth]{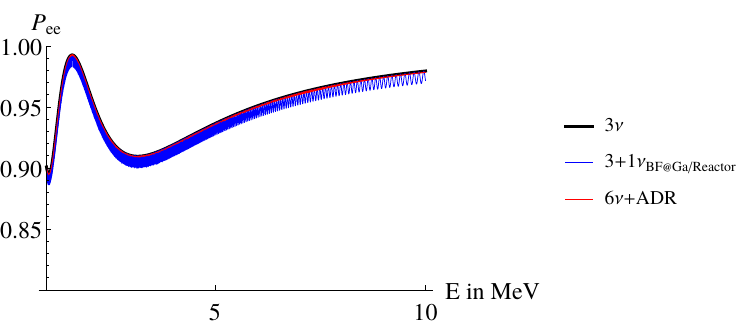}}
	\caption{Disappearance probabilities at NEOS (a), KamLAND (b) and Daya-Bay (c) for BMP 3.
	The blue curve shows the global best fit point for a 3+1-scenario for reactor and gallium experiments found by \cite{Dentler:2018sju} with $\Delta m^2_\text{BF}=1.3\,\text{eV}^2$ and $\sin^2{2\theta_{14}}=0.01$.}
	\label{fig:ReactorsBMP3}
\end{figure}

The oscillation curve at reactor experiments show (see fig. \ref{fig:ReactorsBMP3} that the BMP is well within the bounds put on $|U_{e4}|$ at this energy scale. The oscillation via the additional mass gap in the ADR scenario can be seen best at KamLAND because of its higher baselength and is slightly 'out of tune' because of the different $\Delta m^2$s in the BF and BMP, which characterize the frequency.

\begin{figure}
	\centering
	\subfigure[$P_{\nu_\mu \rightarrow\nu_\mu}$]{\includegraphics[width=0.44\textwidth]{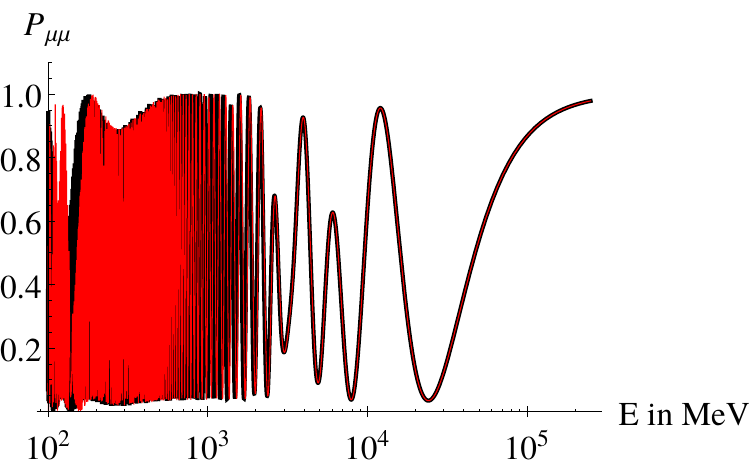}}
	\subfigure[$P_{\nu_\mu \rightarrow\nu_e}$]{\includegraphics[width=0.55\textwidth]{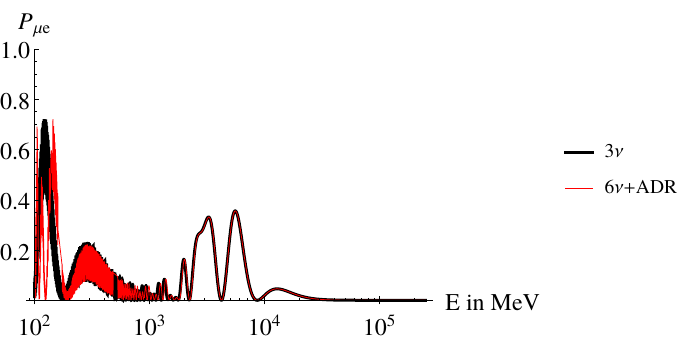}} 
	\caption{Disappearance and appearance probabilities of upward going neutrinos at atmospheric experiments for BMP 3 , where $L \sim d_\text{Earth}$.}
	\label{fig:AtmosphericBMP3}
\end{figure}

\begin{figure}
	\centering
	\subfigure[$P_{\nu_\mu \rightarrow\nu_\mu}$]{\includegraphics[width=0.44\textwidth]{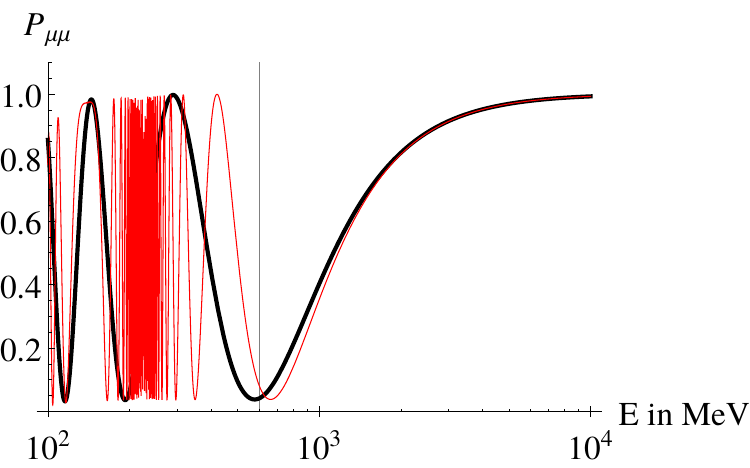}}
	\subfigure[$P_{\nu_\mu \rightarrow\nu_e}$]{\includegraphics[width=0.55\textwidth]{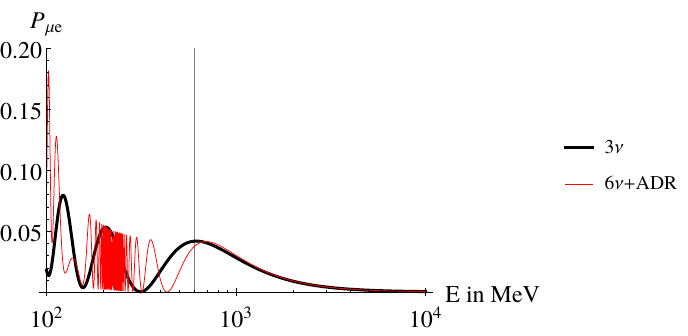}} \\
	\subfigure[$P_{\nu_\mu \rightarrow\nu_\mu}$]{\includegraphics[width=0.44\textwidth]{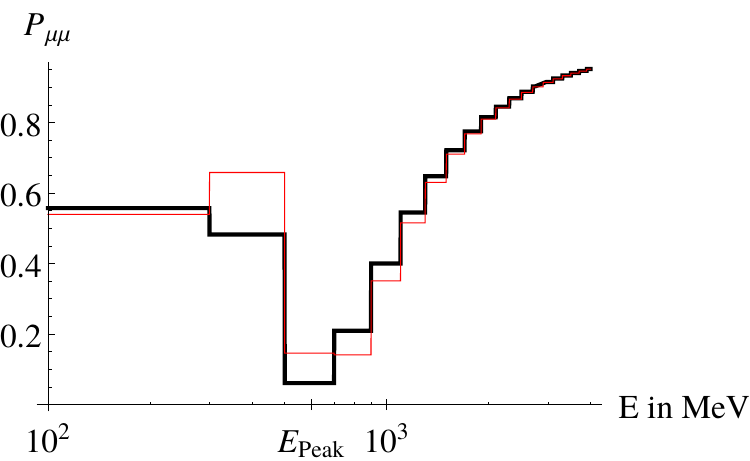}}
	\subfigure[$P_{\nu_\mu \rightarrow\nu_e}$]{\includegraphics[width=0.55\textwidth]{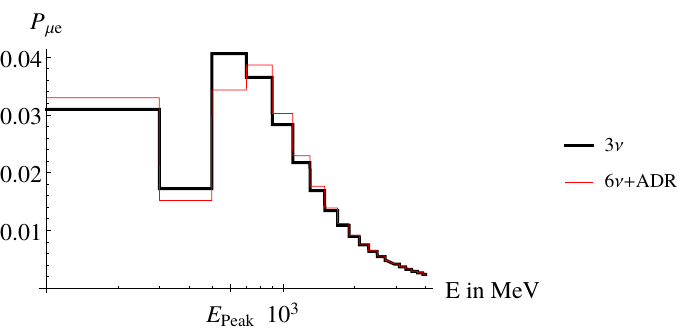}}
	\caption{Disappearance a) and appearance b) probabilities at T2K for BMP 3 and their corresponding binned versions c) and d) with a bin-size of $200\,$MeV. The grey line indicates the peak energy of the accelerators neutrino spectrum.}
	\label{fig:T2KBMP3}
\end{figure}

As in the discussion of BMP2, the upward-going neutrino channel (fig. \ref{fig:AtmosphericBMP3}) at atmospheric experiments is very much indistinguishable from the standard 3$\nu$ paradigm, while averaging effects at T2K can camouflage the effects of the ADR in the transition regime between $200\,$ and $600\,$MeV (fig. \ref{fig:T2KBMP3}). Again, the solar neutrino spectrum is not distorted.

 \FloatBarrier

\subsubsection{Benchmark Point 4}
The BSM parameters of this BMP are 

\begin{align}
	\begin{split}
		\Delta m_\text{LSND}^2 &= 30\, \text{eV}^2 \, ,\sin^2{\theta}=10^{-4}\, ,  \\ 
		\varepsilon= 3 \cdot 10^{-16} \, , \eta &= 100.9 \, , \kappa= 100.7 \, , \xi= 1 \, \\
		\Rightarrow E_{\text{Res},\eta}&=22.3\, \text{MeV} \, , \\
		E_{\text{Res},\kappa}&= 22.3\, \text{MeV} \, , \\
		E_{\text{Res},\xi}&=223.6\, \text{MeV} \,  .
	\end{split}
	\label{eq:BMP4}
\end{align}

\begin{figure}
	\centering
	\includegraphics[width=0.8\textwidth]{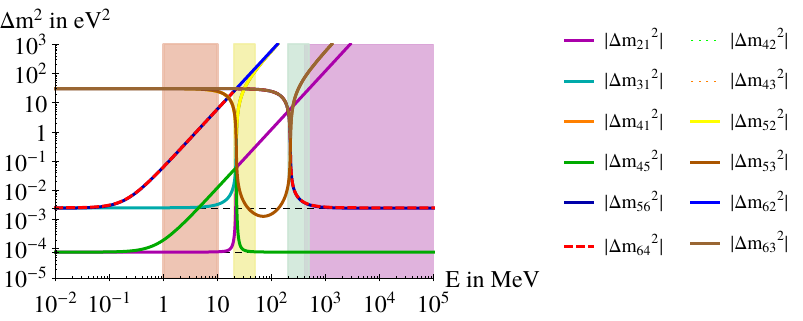}
	\caption{Effective $\Delta m^2$ depending on the energy $E$ for BMP 4.}
	\label{fig:dmSquaredBMP4}
\end{figure}

The corresponding $\Delta m^2(E)$ can be seen in fig. \ref{fig:dmSquaredBMP4}. 

The overview shows that there are level crossings at the desired resonance energies $E_{R,\xi}\sim 25 \,$MeV and $E_{R,\eta / \kappa}\sim 220\,$MeV, meaning in the LSND and MiniBooNE energy ranges respectively. 
It becomes apparent that also in the high $\Delta m^2$ regime an effective 3+1-scenario at energies far below the resonance is recovered. The 3+1$\nu$ modulation has a high frequency in this case because of the high $\Delta m^2_\text{LSND}$, but because of the miniscule mixing angle $\theta$ there is next to no deviation from the $3 \nu$ oscillation probability.
In the long baseline energy range, the relevant (meaning non-decoupled) $\Delta m^2s$ have already relaxed to the standard $\Delta m^2_{21}$ and $\Delta m^2_{31}$, which means that a $3\nu$-scenario is recovered quite well.
 
\begin{figure}
	\centering
	\subfigure[$P_{\nu_\mu\rightarrow\nu_e}$]{\includegraphics[width=0.44\textwidth]{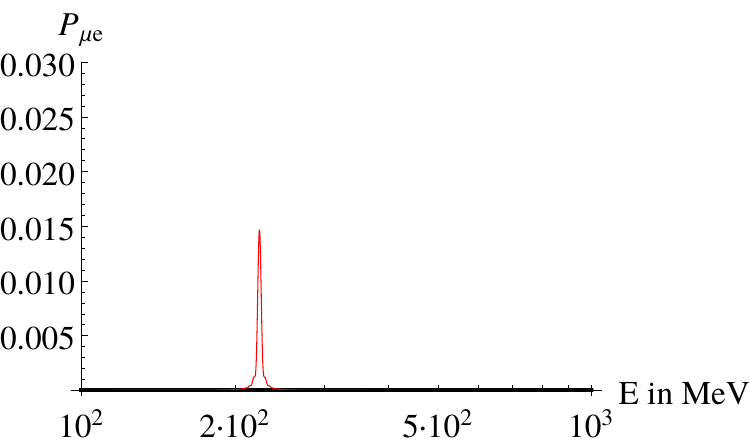}}
	\subfigure[$P_{\nu_\mu\rightarrow\nu_\mu}$]{\includegraphics[width=0.55\textwidth]{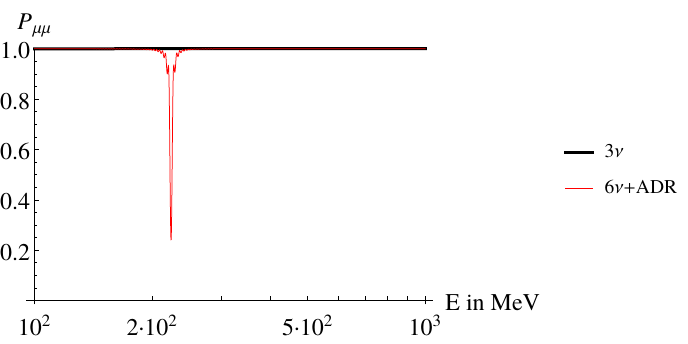}}
	\caption{Appearance (a) and disappearance (b) probabilities at MiniBooNE for BMP 4.}
	\label{fig:MiniBooNEBMP4}
\end{figure}

\begin{figure}
	\centering
	\subfigure[$P_{\nu_\mu\rightarrow\nu_e} @ \text{LSND}$]{\includegraphics[width=0.44\textwidth]{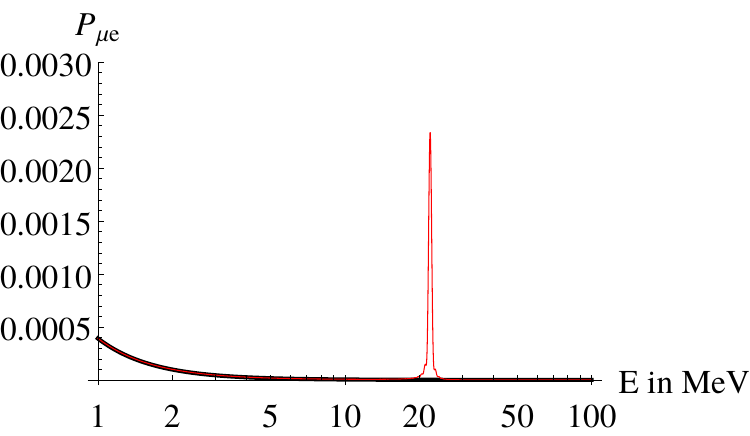}}
	\subfigure[$P_{\nu_\mu\rightarrow\nu_e} @ \text{KARMEN}$]{\includegraphics[width=0.55\textwidth]{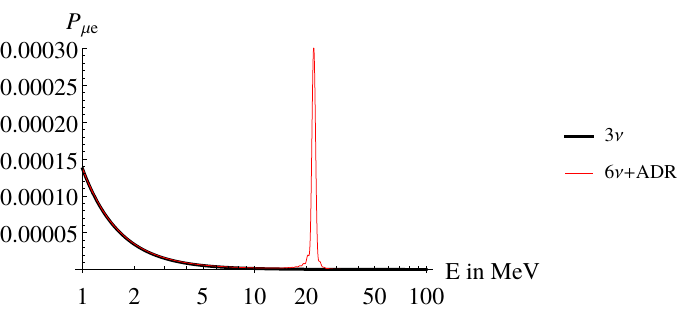}}
	\caption{LSND (a) and KARMEN (b) oscillation probabilities for BMP 4.}
	\label{fig:LSNDKARMENBMP4}
\end{figure}

The oscillation curves for the appearance and disappearance channels at MiniBooNE, as well as the appearance at LSND and KARMEN are plotted in \ref{fig:MiniBooNEBMP4} and \ref{fig:LSNDKARMENBMP4} respectively.

The desired resonance feature is clearly visible in this scenario in the appearance channel at MiniBooNE and LSND at the corresponding resonance energies. The disapperance channel at MiniBooNE also features a sharp dip at the resonance energy, since the conversion to sterile neutrinos is very efficient at that particular energy. KARMEN also features a resonance, but in a much less pronounced way. Because of the tiny mixing, the resonance behavior in this BMP is very distinguished and sharp. 

\begin{figure}
	\centering
	\subfigure[$P_{\nu_e \rightarrow\nu_e} @ \text{NEOS}$]{\includegraphics[width=0.49\textwidth]{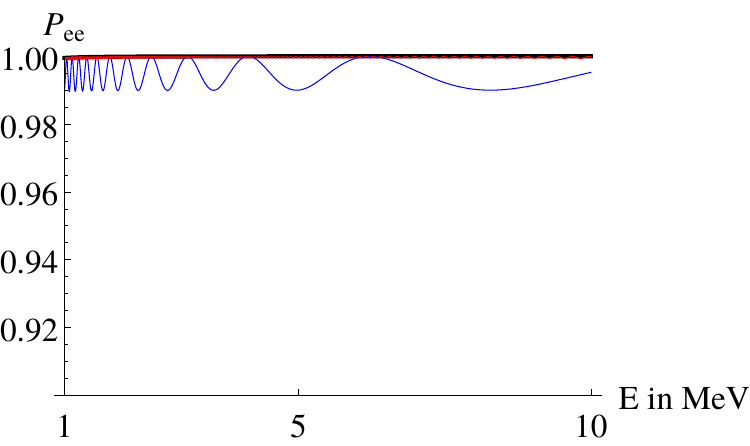}}
	\subfigure[$P_{\nu_e \rightarrow\nu_e} @ \text{KamLAND}$]{\includegraphics[width=0.49\textwidth]{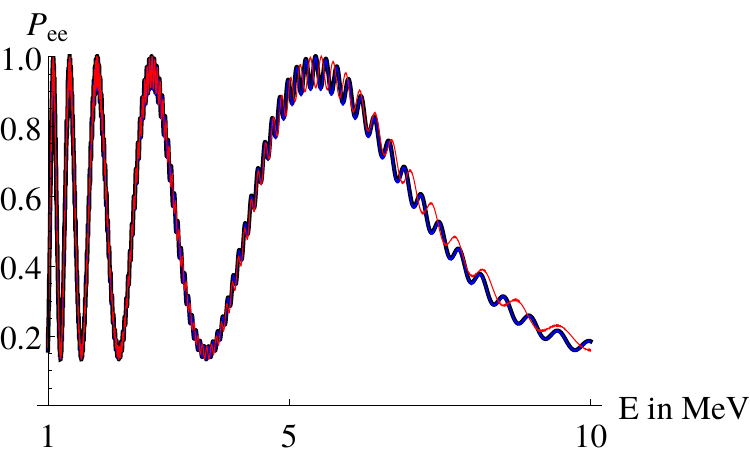}} \\
	\subfigure[$P_{\nu_e \rightarrow\nu_e} @ \text{Daya-Bay}$]{\includegraphics[width=0.63\textwidth]{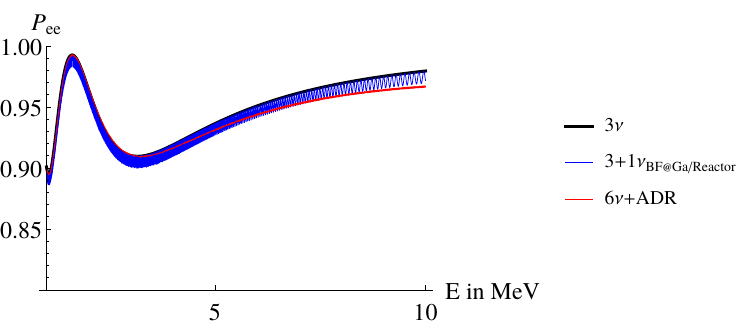}}
	\caption{Disappearance probabilities at NEOS (a), KamLAND (b) and Daya-Bay (c) for BMP 4.
	The blue curve shows the global best fit point for a 3+1-scenario for reactor and gallium experiments found by \cite{Dentler:2018sju} with $\Delta m^2_\text{BF}=1.3\,\text{eV}^2$ and $\sin^2{2\theta_{14}}=0.01$.}
	\label{fig:ReactorsBMP4}
\end{figure}

We plotted the oscillation curves for the low energy reactor experiments in fig. \ref{fig:ReactorsBMP4}.

As suggested in the discussion above and in section \ref{sec:PhenoBelowResonance}, the oscillation has already almost fully relaxed to an effective 3+1 scenario. The 'out-of-tune' effect at KamLAND, which was observable at BMP3 is again visible in this case, since the $\Delta m^2$ is the same. Because of the small mixing angle $\theta$ there is almost no visible oscillation in the short baseline experiments Daya-Bay and NEOS. Interestingly, at Daya-Bay, the ADR curve almost follows the averaging line of the 3+1-BF-scenario, meaning it could potentially mimic a 3+1-scenario in the data. 

\begin{figure}
	\centering
	\subfigure[$P_{\nu_\mu \rightarrow\nu_\mu}$]{\includegraphics[width=0.44\textwidth]{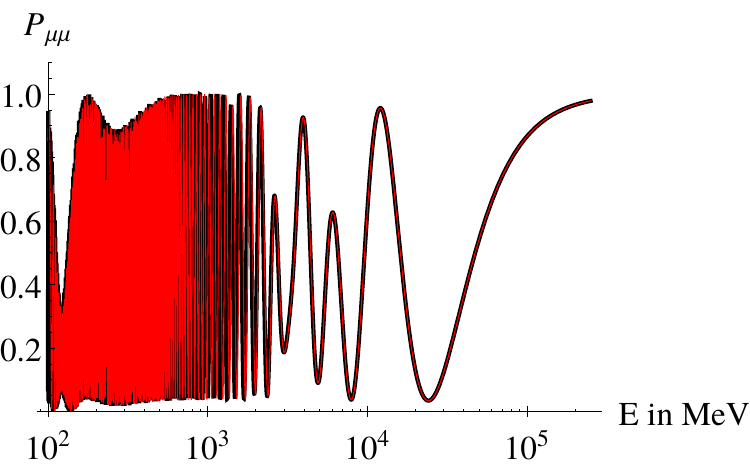}}
	\subfigure[$P_{\nu_\mu \rightarrow\nu_e}$]{\includegraphics[width=0.55\textwidth]{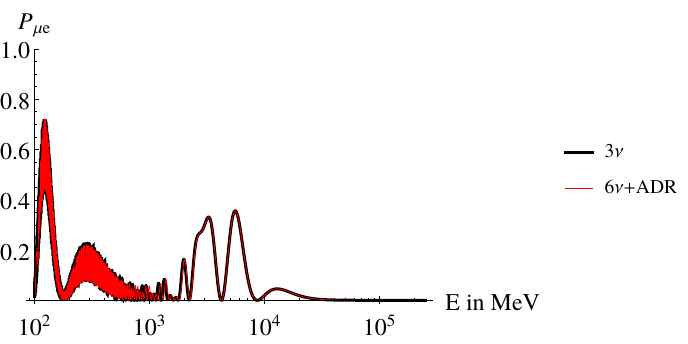}} 
	\caption{Disappearance and appearance probabilities of upward going neutrinos at atmospheric experiments for BMP 4 , where $L \sim d_\text{Earth}$.}
	\label{fig:AtmosphericBMP4}
\end{figure}

One can see on the oscillation probability level already, that there is no possibility to distinguish BMP3 from the 3$\nu$ scenario in the upward-going neutrino channel (fig. \ref{fig:AtmosphericBMP4}), since the coherent forward scattering within earth and the ADR potential have the same sign and therefore work in the same direction, but with different energy dependences. This makes convergence to the measured eigenvalues even faster and therefore washes out any detectable effects on the probability.    

\begin{figure}
	\centering
	\subfigure[$P_{\nu_\mu \rightarrow\nu_\mu}$]{\includegraphics[width=0.44\textwidth]{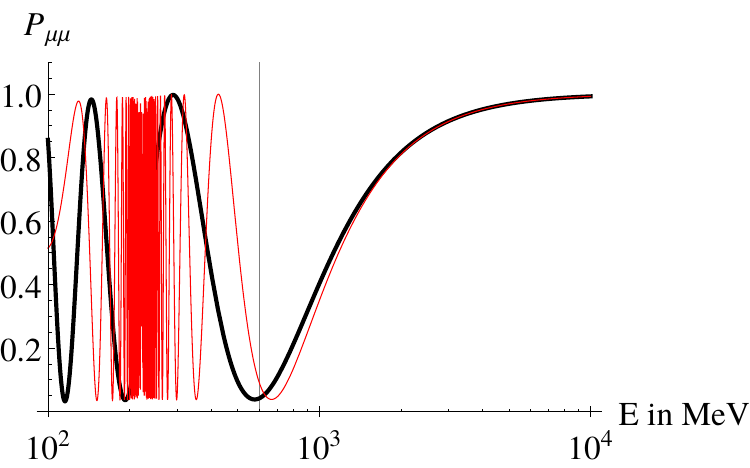}}
	\subfigure[$P_{\nu_\mu \rightarrow\nu_e}$]{\includegraphics[width=0.55\textwidth]{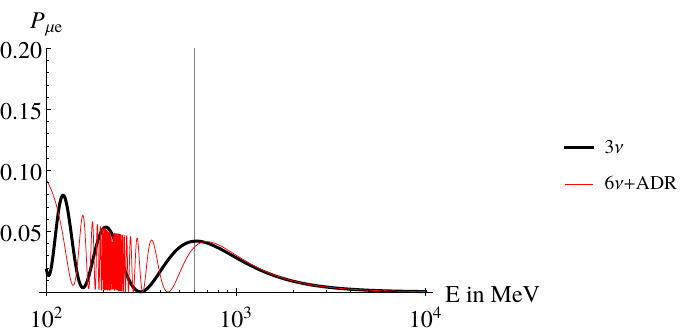}} \\
	\subfigure[$P_{\nu_\mu \rightarrow\nu_\mu}$]{\includegraphics[width=0.44\textwidth]{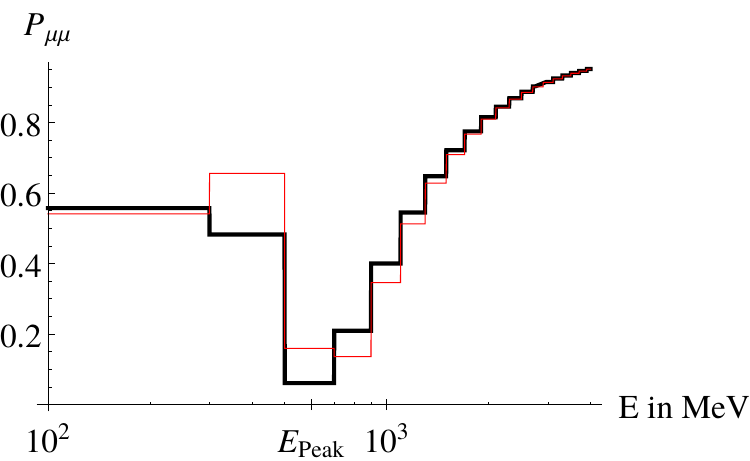}}
	\subfigure[$P_{\nu_\mu \rightarrow\nu_e}$]{\includegraphics[width=0.55\textwidth]{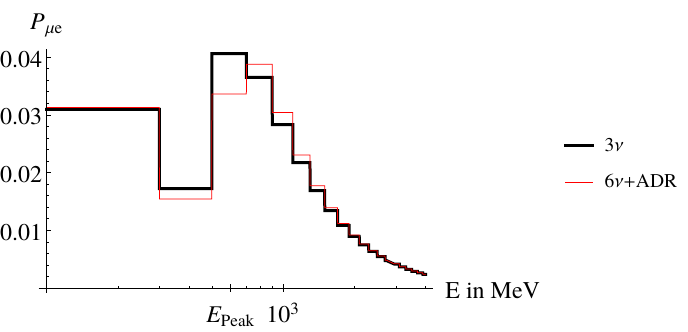}}
	\caption{Disappearance a) and appearance b) probabilities at T2K for BMP 4 and their corresponding binned versions c) and d) with a bin-size of $200\,$MeV. The grey line indicates the peak energy of the accelerators neutrino spectrum.}
	\label{fig:T2KBMP4}
\end{figure}

In fig. \ref{fig:T2KBMP4}, we show the oscillation curves and bin plots for T2K. The effect is the same as in BMP 2 and BMP 3 in this case.

For this parameter combination the ADR only starts to affect the solar $\Delta m^2_{21}$ around $20\,$MeV when the first level crossings occur which is above the maximum energy of ${}^{8}$B neutrinos and far above the peak of the energy spectrum. Thus no significant distortion of solar neutrino spectra is expected for BMP 4.

 \FloatBarrier

\section{Open Questions}\label{sec:openquestions}
\begin{figure}
	\centering
	\subfigure[Disappearance $@$ BMP3]{\includegraphics[width=0.44\textwidth]{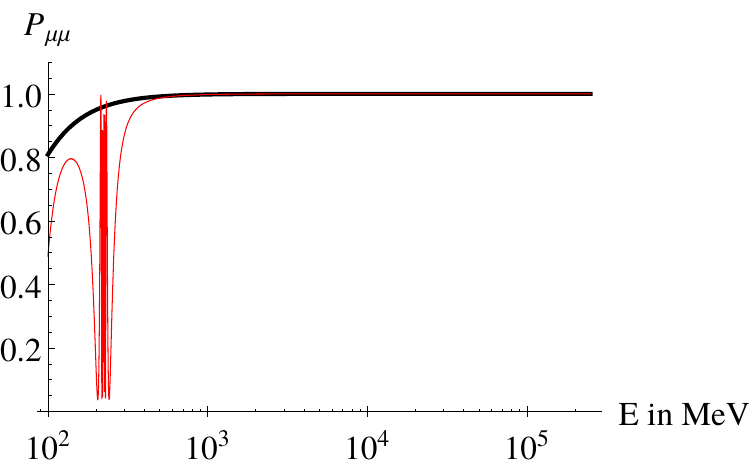}}
	\subfigure[Appearance $@$ BMP3]{\includegraphics[width=0.55\textwidth]{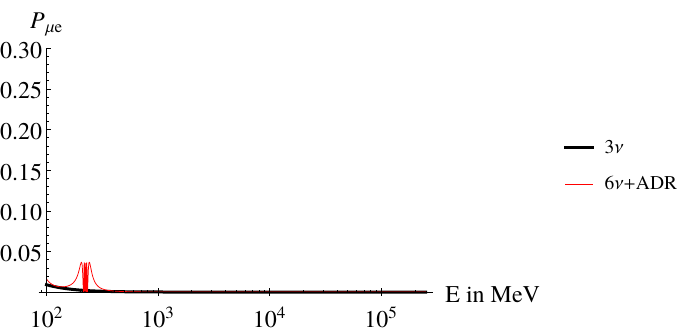}}\\
	\subfigure[Disappearance $@$ BMP4]{\includegraphics[width=0.44\textwidth]{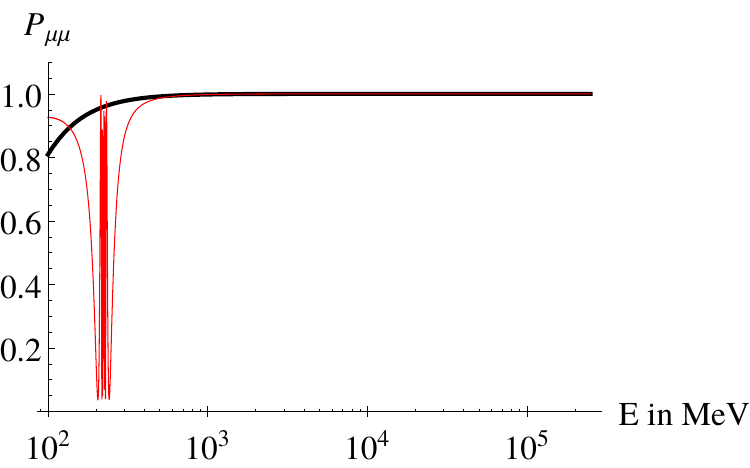}}
	\subfigure[Appearance $@$ BMP4]{\includegraphics[width=0.55\textwidth]{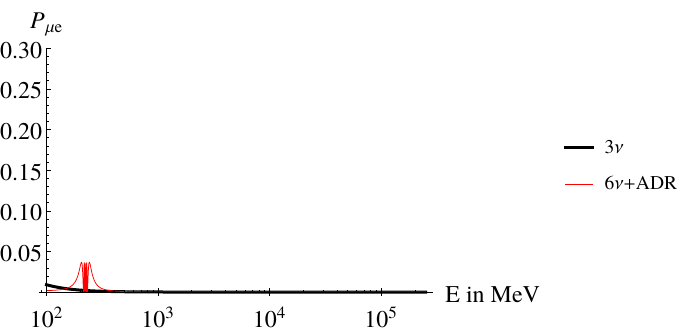}}\\
	\caption{Different Probabilities for downward-going neutrinos at atmospheric neutrino experiments that highlight the Sub-GeV to $\sim 100\,$GeV region.}
	\label{fig:DownwardGoing}
\end{figure}

\begin{figure}
	\centering
	\subfigure[Disappearance $@$ BMP3 with bin]{\includegraphics[width=0.44\textwidth]{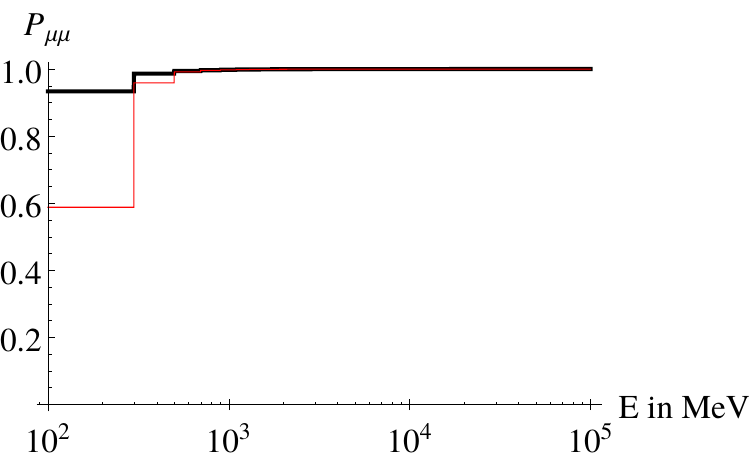}}
	\subfigure[Appearance $@$ BMP3 with bins]{\includegraphics[width=0.55\textwidth]{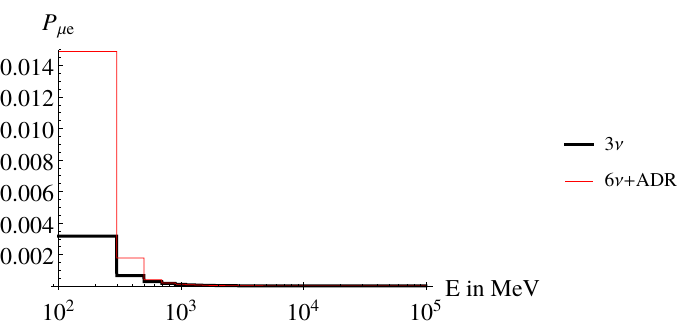}}\\
	\subfigure[Disappearance $@$ BMP4 with bins]{\includegraphics[width=0.44\textwidth]{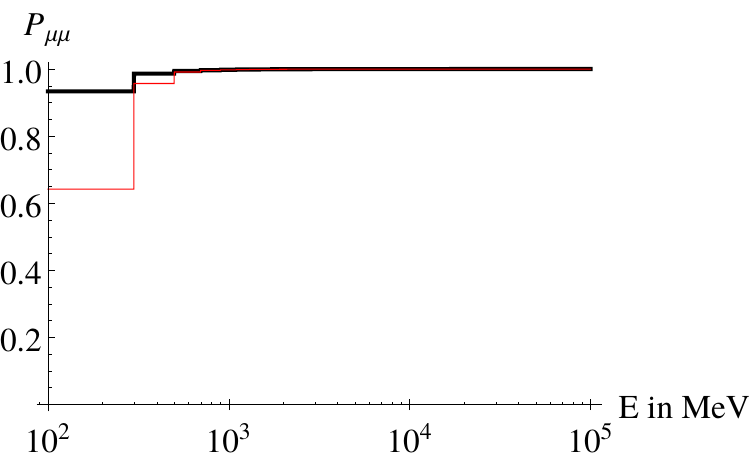}}
	\subfigure[Appearance $@$ BMP4 with bins]{\includegraphics[width=0.55\textwidth]{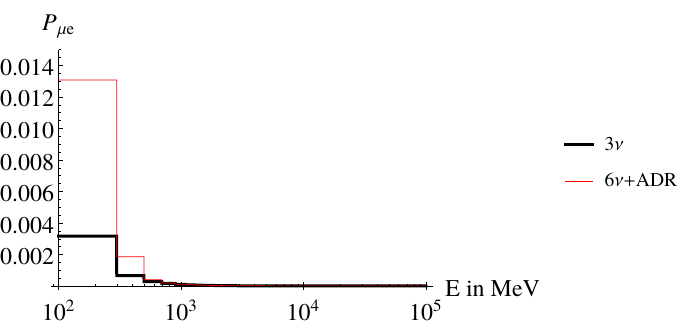}}\\
	\caption{Different Probabilities for downward-going neutrinos at atmospheric neutrino experiments with finite energy resolution. Bin size is $200\,$MeV.}
	\label{fig:DownwardGoingBins}
\end{figure}

As can be seen in sec. \ref{sec:Pheno}, the proposed model resembles three neutrino oscillations far above
the resonance (in the GeV region). However, due to the desired resonance at around $200\,$MeV for explaining
the MiniBooNE data, this resonance does also have impact on the sub-GeV neutrinos at LBL experiments. While we show that the upward-going neutrino channels are reasonably safe, due to the relatively strong matter effects within the earth that push the $\Delta m^2$s towards faster convergence, the downward-going neutrinos do not experience such effects.
The corresponding oscillations in this energy region are plotted in Fig. \ref{fig:DownwardGoing} for two exemplary BMPs . As expected, the model predicts a deviation from the simple three neutrino model in the transition area. Nevertheless, without access to the actual data it is hard to judge whether these oscillation patterns are excluded by current experiments.
Previous analyses~\cite{TheIceCube:2016oqi, Abe:2014gda} searched for sterile neutrinos without altered dispersion relations.
Since the plots again feature fast oscillations, that cannot be seen by any experiment because of the finite energy resolution, we included a binned version of Fig. \ref{fig:DownwardGoing} in Fig. \ref{fig:DownwardGoingBins}. While there exists an obvious deviation from the predictions of the proposed scenario, the proposed exclusion limits in the literature cannot be adopted for our case. Furthermore, the true neutrino energy spectra from recent analyses (e.g. \cite{Abe:2017aap}) indicate that there is almost no statistics in the energy range below $200\,$MeV at Super Kamiokande. Additonally, a recent IceCube analysis \cite{Aartsen:2020iky} found hints for an additional oscillation over the $\Delta m^2 \sim 8\,\text{eV}^2$ mass gap, which is a slight indication of deviation from the 3$\nu$ picture.
To either exclude or confirm the model proposed in this article, we recommend a reanalysis of the current sub-GeV data in atmospheric experiments.

\section{Cosmological bounds}

Apart from neutrino oscillation experiments, cosmology provides stringent bounds on any early-Universe population of
eV-scale sterile neutrinos.
Such neutrinos constitute additional radiation degrees of freedom which alter the Hubble expansion and consequently via 
the freeze-out temperature of electroweak interactions and the proton-to neutron ratio the cosmic abundances
of light elements explained very well in Big Bang nucleosynthesis (BBN). 
Moreover, eV~scale neutrinos are way above the cosmological neutrino mass bounds due to the effect that
such heavy neutrinos efficiently suppress the formation of structure on scales smaller than their free-streaming length.
As proposed in ~\cite{Pas:2005rb} and analyzed in detail in~\cite{Aeikens:2016rep, Jang:2016rpi}, 
altered dispersion relations suppress active-sterile neutrino mixing above the resonance and thus
can prevent sterile neutrinos from being populated at high energies. While naively the resonance energies discussed in this work appear 
to be too high to prevent sterile neutrinos from being populated at the MeV scale relevant for BBN and neutrino decoupling, 
it is conceivable that the responsible Lorentz violating altered dispersion relation depends on temperature and density.
This occurs for example in extra-dimensional ADR models where Einstein's equations obtain new source terms due to the hot dense plasma
in the early universe, 
which leads to modifications in the metric of the warped extra dimensions and therefore again alters the dispersion relation.
Such considerations are strongly model dependent, and are beyond the scope of this paper. 
It should be stressed, however, that altered dispersion relations proposed here as a solution for the MiniBooNE and LSND anomalies 
may also help to evade the cosmological bounds usually applied to light sterile neutrinos.

\section{Summary and Outlook}
In this paper we have developed a 3+3 neutrino framework with altered dispersion relations, 
that can explain different combinations of the LSND, MiniBooNE, Reactor and Gallium anomalies.
We have presented different benchmark points that elucidate the parameter space and characteristic features of the model.
It has been shown that the model can accomodate two resonances, one in the oscillation amplitude resembling the low-energy excess in the MiniBooNE data, and one at lower energy to enhance the LSND signal. While not all anomalies can be explained at once, giving up either the Ga-/Reactor-, LSND- or MiniBooNE anomaly can complete the rest of the picture.
To our best knowledge, this is the only model which can achieve the task at hand to this extent. 
We have discussed in detail which constructions fail in this context, 
and how we arrived at the scenario which seems to provide the most promising explanation of the data. 

To arrive at this conclusion, we first have pointed out that for 3+1 neutrino models with sterile neutrino altered dispersion relations, the predominantly sterile state decouples 
at energies far above the resonance, which thus hides the sterile neutrino in accelerator experiments operating at high energies \cite{Chung:1999xg, Csaki:2000dm}.
As has been recently pointed out \cite{Doring:2018ncz}, such a scenario results as the effective field theory limit of sterile neutrinos 
propagating in an asymmetrically warped extra dimension. 
However, a level crossing occurs at the resonance energy, and the Hamiltonian eigenstates swap their flavor content. 
While the predominantly sterile state decouples above the resonance as discussed above, 
the now heavier predominantly active state approaches a constant value that is different from its initial value because of the level crossing gap. 
This implies a large effective $\Delta m^2_{13}$ that gives rise to large and fast active-to-active oscillations
(e.g.~in atmospheric neutrinos) which rules out the 3+1$\nu$ model.

This consequence can be avoided when 3 active and 3 sterile neutrinos are introduced and the sterile neutrinos are mixed with the active ones
via a common effective potential.   In such a model, $E_R$ is 
necessarily common too.
In this case all Hamiltonian states are altered in the same way by the admixture with the ADR-influenced sterile neutrino 
(they change in parallel and the effective $\Delta m^2$'s 
remain constant). While this feature solves the problem above, it also
leads also to a cancellation of active-to-active oscillations over the LSND mass gap due to unitarity constraints - and thus implies there are no oscillations at all at 
both LSND and MiniBooNE, as long as all sterile neutrinos feature the same resonance energy.

However, in the scenario with three active and three sterile neutrinos, the high energy limit of the Hamiltonian eigenvalues for the 
predominantly active neutrinos is independent of both energy and the ADR parameter parametrizing the altered dispersion relation. 
So once one assigns different resonance energies 
 to the 3 sterile neutrinos,  
it is possible to obtain oscillations and resonances in the intermediate mass regime while an
effective 3-active neutrino scenario is restored at high energies.  
New parameters are inevitably introduced via the sterile neutrino sector.  

While it is possible that the various neutrino anomalies are due to our limited understanding of experimental backgrounds,
this ``Beyond the Standard Model'' physics scenario is testable by the MicroBooNE and ICARUS experiments, 
and may reveal itself first in sub-GeV atmospheric neutrino data.

\newpage
 
\textit{Note Added:} While the first version of this paper was under revision, the article \cite{Barenboim:2019hso} appeared, which highlighted the constraints on this model by studying two exemplary parameter sets. While we agree with the significance of the constraints discussed for the given data points, we disagree with the general conclusion that studying two points in a five-dimensional parameter space is sufficient to prove it incompatible with global data.

\bibliographystyle{ksfhnat}

\end{document}